\documentclass[aps,prx,twocolumn,amsmath,amssymb,floatfix,nofootinbib,superscriptaddress]{revtex4-2}
\usepackage{graphicx} 
\usepackage{bm} 
\usepackage{hyperref} 

\usepackage{braket,bbold,color,enumerate,makecell,multirow}

\usepackage{amsmath, amsfonts, amssymb, mathtools}
\usepackage{xcolor}

\newcommand{\ii}{\text{i}}
\newcommand{\CS}{\mathcal{C}}
\newcommand{\bs}[1]{\boldsymbol{#1}}

\begin{document}

\title{Hermitian Bulk -- Non-Hermitian Boundary Correspondence}

\author{Frank Schindler}
\affiliation{Princeton Center for Theoretical Science, Princeton University, Princeton, NJ 08544, USA}
\affiliation{Blackett Laboratory, Imperial College London, London SW7 2AZ, United Kingdom}

\author{Kaiyuan Gu}
\affiliation{Department of Physics, Princeton University, Princeton, NJ 08544, USA}

\author{Biao Lian}
\affiliation{Department of Physics, Princeton University, Princeton, NJ 08544, USA}

\author{Kohei Kawabata}
\affiliation{Department of Physics, Princeton University, Princeton, NJ 08544, USA}
\affiliation{Institute for Solid State Physics, University of Tokyo, Kashiwa, Chiba 277-8581, Japan}

\begin{abstract}
Non-Hermitian band theory distinguishes between line gaps and point gaps. While point gaps can give rise to intrinsic non-Hermitian band topology without Hermitian counterparts, line-gapped systems can always be adiabatically deformed to a Hermitian limit. Here we show that line-gap topology and point-gap topology can be intricately connected: topological line-gapped systems in $d$ dimensions induce nontrivial point-gap topology \emph{on their $(d-1)$-dimensional boundaries} when suitable internal and spatial symmetries are present. Since line-gapped systems essentially realize Hermitian topological phases, this establishes a correspondence between Hermitian bulk topology and intrinsic non-Hermitian boundary topology. For the correspondence to hold, no non-Hermitian perturbations are required in the bulk itself, so that the bulk can be purely Hermitian. Concomitantly, the presence of non-Hermitian perturbations in the bulk does not affect any results as long as they do not close the bulk line gap. On the other hand, non-Hermitian perturbations are essential on the boundary to open a point gap. The non-Hermitian boundary topology then further leads to higher-order skin modes, as well as chiral and helical hinge modes, that are protected by point gaps and hence unique to non-Hermitian systems. We identify all the internal symmetry classes where bulk line-gap topology induces boundary point-gap topology as long as an additional spatial symmetry is present, and establish the correspondence between their topological invariants. There also exist some symmetry classes where the Hermitian edge states remain stable, in the sense that even a point gap cannot open on the boundary.
\end{abstract}

\maketitle

\section{Introduction}
Hermitian topological insulators and superconductors in the ten Altland-Zirnbauer (AZ) symmetry classes~\cite{AZ-97} are gapped in their $d$-dimensional bulk but exhibit gapless edge states on $(d-1)$-dimensional boundaries~\cite{Hasan-Kane-review, Qi-Zhang-review, Schnyder-Ryu-review}. These edge states are protected by the bulk gap and symmetries and cannot be removed by local symmetry-preserving perturbations without a phase transition. In non-Hermitian (NH) systems, two notions of gap, line gaps and point gaps, are fundamental for topological phases~\cite{Shen2018, Gong2018, Kawabata2019}.
While line-gapped systems are always deformable to either a Hermitian (real line gap) or anti-Hermitian (imaginary line gap) gapped limit, point-gapped systems may not have any gapped Hermitian or anti-Hermitian counterparts. Correspondingly, point-gapped systems, as well as line-gapped systems that only have an anti-Hermitian limit, may realize a radically different kind of topology that is intrinsic to NH systems. In the following, we use the term \emph{NH topology} for these new topological equivalence classes that arise when a point gap or imaginary line gap is kept open. At the same time, we refer to real line gap topology as \emph{Hermitian topology}, as it can be obtained by perturbing around gapped Hermitian systems.
Similar to Hermitian topological phases, nontrivial NH topology in $d$ dimensions is reflected by a gap closing on $(d-1)$-dimensional boundaries. In many cases, this gap closing induces a NH skin effect~\cite{ZhangYangFang2020, Okuma2020_toposkin}, where a macroscopic number of eigenstates are localized at the system boundaries~\cite{Lee2016, Torres2018, Wang2018, Kunst2018, Lee2019_skineffect, Yokomizo_Murakami_2019,Borgnia2020}. 
Some NH systems in two dimensions (2D) and three dimensions (3D) may furthermore avoid a skin effect on (almost all) boundaries and instead exhibit anomalous NH boundary dispersion such as unpaired exceptional points~\cite{TerrierKunst_2020, eti, Kawabata2021, Nakamura2022, Denner23}.

Here, we uncover an intricate connection between Hermitian topology in $d$ dimensions and NH topology on their $(d-1)$-dimensional boundaries that is guaranteed by spatial symmetry. 
We consider Hermitian topological insulators and superconductors in the presence of small NH perturbations. The perturbations are chosen such that they preserve a given NH symmetry which reduces to the appropriate AZ symmetry in the Hermitian limit. 
This requirement ensures that the edge states of the original phase cannot acquire a real line gap. Conversely, the perturbations generically open either a point gap or an imaginary line gap on the boundary. In a geometry with open boundary conditions (OBC) in one direction and periodic boundary conditions (PBC) in the $(d-1)$ remaining directions (the ``slab geometry”), the NH perturbed system can therefore in principle realize the intrinsic NH topological phase of a $(d-1)$-dimensional bulk NH system.
We solidify this possibility by proving that in the presence of pseudo-inversion symmetry (a simple spatial symmetry that is compatible with nontrivial NH topology~\cite{Okugawa2021, Vecsei2021, ShiozakiOno2021}), nontrivial NH topology is \emph{guaranteed} on the boundaries of certain Hermitian topological phases. 
This statement holds irrespective of the choice of specific model parameters or boundary terminations, and only relies on the presence of a point gap in the slab geometry. Thus, while the mere opening of a point gap does not necessarily lead to NH topology, we demonstrate that the combination of a point gap and Hermitian topology in the bulk must give rise to boundary NH topology, as long as pseudo-inversion symmetry is preserved.
We derive the resulting mapping between NH boundary topological invariants and Hermitian bulk invariants in all AZ symmetry classes and their NH extensions (Tabs.~\ref{tab: mapping2D} and \ref{tab: mapping3D}). We also show that $d$-dimensional Hermitian topological phases in the presence of small NH perturbations host, via their boundary-induced $(d-1)$-dimensional NH topology, $(d-2)$-dimensional \emph{higher-order} NH gapless states such as higher-order skin modes and chiral or hinge modes protected by a point gap and pseudo-inversion symmetry.
This phenomenon constitutes a novel kind of Hermitian-NH hybrid higher-order topology. 
Unlike previous Hermitian and NH examples of higher-order corner and hinge states~\cite{Tao2019, Lee2019, Evardsson2019, Zhang2019, Luo2019, Okugawa2020, Kawabata2020, FuHuWan2021, Kim2021}, the higher-order skin modes arising from this mechanism do not appear on opposite boundaries.

\begin{table}[t]
\caption{\label{tab: mapping2D}Correspondence between 2D bulk line-gap topology and 1D boundary point-gap topology. 
For each Hermitian symmetry class (H class), we show the topological classification for the 2D bulk in parentheses, and the corresponding topological invariant in the second column (Chern number $C \in \mathbb{Z}$ and $\mathbb{Z}_2$ invariant $\nu \in \{0,1\}$). 
For each NH symmetry class, we show the point-gap topological classification~\cite{Kawabata2019} for the 1D boundaries in parentheses, and the corresponding topological invariant in the fourth column [complex-spectral winding number $W(E) \in \mathbb{Z}$ and its $\mathbb{Z}_2$ analog $\nu(E) \in \{ 0,1 \}$, both of which depend on a reference energy $E$ in the complex plane]. 
The boundary NH invariants are expressed in terms of the respective bulk Hermitian invariants for a slab geometry preserving pseudo-inversion symmetry $\mathcal{I}^\dagger$ [Eq.~\eqref{eq: pseudo inversion}]. ``--" indicates that there is no correspondence.
}
\begin{tabular}{c|c||c|c} \hline \hline
\multicolumn{2}{c||}{\thead{$d=2$ \\ line-gap topology}} & \multicolumn{2}{c}{\thead{$d=1$ \\ point-gap topology}} \\ \hline
H class & invariant & NH class & invariant (with $\mathcal{I}^\dagger$) \\ \hline \hline
A ($\mathbb{Z}$) & $C \in \mathbb{Z}$ & A ($\mathbb{Z}$) & $W(E_\mathrm{F}) = C$ (mod $2$) \\ \hline
\multirow{2}{*}{D ($\mathbb{Z}$)} & \multirow{2}{*}{$C \in \mathbb{Z}$} & D ($\mathbb{Z}_2$) & -- \\ 
    &                   & D$^\dagger$ ($\mathbb{Z}$) & $W(0) = C$ (mod $2$) \\ \hline
DIII ($\mathbb{Z}_2$) & $\nu \in \{0,1\}$ & DIII$^\dagger$ ($\mathbb{Z}_2$) & $\nu(0) = \nu$ \\ \hline
\multirow{2}{*}{AII ($\mathbb{Z}_2$)} & \multirow{2}{*}{$\nu \in \{0,1\}$} & AII ($2\mathbb{Z}$) & $W(E_\mathrm{F}) = 2\nu$ (mod $4$) \\
    &                   & AII$^\dagger$ ($\mathbb{Z}_2$) & $\nu(E_\mathrm{F}) = \nu$\\ \hline
C (2$\mathbb{Z}$) & $C \in 2\mathbb{Z}$ & C$^\dagger$ (2$\mathbb{Z}$) & $W(0) = C$ (mod $4$) \\  \hline \hline
\end{tabular}
\end{table}

We note that Ref.~\onlinecite{ou2022non} recently discussed NH topological phases at the boundaries of 2D systems. However, the nontrivial point-gap topology discussed in Ref.~\onlinecite{ou2022non} does not rely on the presence of a 2D bulk but can instead be viewed as that of an independent one-dimensional (1D) NH system. 
This is not the case for our work, where each boundary carries an anomalous \emph{half}~\cite{Schindler2022} of a $(d-1)$-dimensional NH system that is stabilized by the presence of a $d$-dimensional bulk with Hermitian topology. Furthermore, Refs.~\onlinecite{LiLiu2022, ZhuGong2022, ZhuGong2023} recently studied the corner skin effect of NH Chern insulators.
However, Refs.~\onlinecite{LiLiu2022, ZhuGong2022, ZhuGong2023} did not find a connection between the Hermitian topology in the 2D bulk and the NH topology in the 1D boundaries.
Here, we clarify this connection and further generalize it to arbitrary AZ symmetry classes and spatial dimensions.
Notably, the NH topology in 2D surfaces can lead to chiral and helical hinges modes protected by point-gap topology, rather than the skin modes.
We classify the correspondence between Hermitian and NH topological invariants that is guaranteed by pseudo-inversion symmetry, as summarized in Table~\ref{tab: mapping2D} for 2D and Table~\ref{tab: mapping3D} for 3D.

\begin{table}[t]
\caption{\label{tab: mapping3D}Correspondence between 3D bulk line-gap topology and 2D boundary point-gap topology. For each Hermitian symmetry class (H class), we show the topological classification for the 3D bulk in parentheses, and the corresponding topological invariant in the second column (winding number $W \in \mathbb{Z}$ and $\mathbb{Z}_2$ invariant $\nu \in \{0,1\}$). For each NH symmetry class, we show the point-gap topological classification~\cite{Kawabata2019} for the 2D boundaries in parentheses, and the corresponding topological invariant in the fourth column [point-gap invariant $C(E) \in \mathbb{Z}$ and its $\mathbb{Z}_2$ analog $\nu(E) \in \{ 0,1 \}$, both of which depend on a reference energy $E$ in the complex plane]. The boundary NH invariants are expressed in terms of the respective bulk Hermitian invariants for a slab geometry preserving pseudo-inversion symmetry $\mathcal{I}^\dagger$ [Eq.~\eqref{eq: pseudo inversion}]. ``--" indicates that there is no correspondence.}
\begin{tabular}{c|c||c|c} \hline \hline
\multicolumn{2}{c||}{\thead{$d=3$ \\ line-gap topology}} & \multicolumn{2}{c}{\thead{$d=2$ \\ point-gap topology}} \\ \hline 
H class & invariant & NH class & invariant (with $\mathcal{I}^\dagger$) \\ \hline \hline
AIII ($\mathbb{Z}$) & $W \in \mathbb{Z}$ & AIII ($\mathbb{Z}$)  & $C(0) = W$ (mod $2$)  \\ \hline
\multirow{2}{*}{DIII ($\mathbb{Z}$)} & \multirow{2}{*}{$W \in \mathbb{Z}$} & DIII ($\mathbb{Z}_2$) & -- \\ 
        &           & DIII$^\dagger$ ($\mathbb{Z}$) & $C(0) = W$ (mod $2$) \\ \hline
AII ($\mathbb{Z}_2$) & $\nu \in \{0,1\}$ & AII$^\dagger$ ($\mathbb{Z}_2$) & $\nu(E_\mathrm{F}) = \nu$ \\ \hline
\multirow{2}{*}{CII ($\mathbb{Z}_2$)} & \multirow{2}{*}{$\nu \in \{0,1\}$} & CII ($2\mathbb{Z}$) & $C(0) = 2\nu$ (mod $4$) \\ 
        &           & CII$^\dagger$ ($\mathbb{Z}_2$) & $\nu(0) = \nu$ \\ \hline
CI ($2\mathbb{Z}$) & $W \in 2\mathbb{Z}$ & CI$^\dagger$ ($2\mathbb{Z}$) & $C(0) = W$ (mod $4$)  \\ \hline \hline
\end{tabular}
\end{table}

\section{2D Chern insulator (class A)} 
    \label{sec: Chern_story}
We begin with the simplest case: Hermitian AZ class A (no symmetry) in 2D. 
This class of systems is characterized by an integer-valued topological invariant, given by the Chern number $C \in \mathbb{Z}$.
As a consequence of this Chern number in the bulk, chiral edge modes appear under OBC.
Below, for both continuum and lattice systems, we show that these chiral edge modes exhibit point-gap topology in the presence of certain NH perturbations, leading to the second-order skin effect at corners (Fig.~\ref{fig: chern_schematic}).

\subsection{Continuum theory}

The elementary topological phase in class A has $|C| = 1$. In a Hermitian Chern insulator under PBC in the $x$-direction and OBC in the $y$-direction (the ``slab geometry"), each boundary hosts a single gapless chiral mode with dispersion $E (k_x) = \pm v k_x$, where $v$ is the Fermi velocity and the sign choice $+$ ($-$) applies to the top (bottom) edge. The low-energy continuum Hamiltonian for the full slab geometry is then given by
\begin{equation}
H_0(k_x) = v k_x \tau_z,
\end{equation}
where we use $2 \times 2$ Pauli matrices $\tau_i$ ($i=0,x,y,z$) such that $\tau_z = +1$ $(-1)$ corresponds to the top (bottom) edge. We assume that the two edges are sufficiently separated so that terms multiplying $\tau_x$ and $\tau_y$ in the Hamiltonian are forbidden by the requirement of locality. We also fix the Fermi level to lie at $E_\mathrm{F} = 0$ for simplicity.

We next add local NH perturbations. In general, we obtain the perturbed Hamiltonian
\begin{equation} \label{eq: perturbedChernSlabHam}
H(k_x) = \mathrm{i} a \tau_0 + (v k_x + \mathrm{i} b) \tau_z,
\end{equation}
with $a,b \in \mathbb{R}$.
Depending on the values of $a$ and $b$, this Hamiltonian may or may not have a nontrivial spectral winding number~\cite{Gong2018, Kawabata2019}
\begin{align}
    W(E) = \int_{-\infty}^{\infty} \frac{dk_x}{2\pi\ii} \frac{d}{dk_x} \log \det \left[ H \left( k_x \right) - E \right],
        \label{eq: 1D winding}
\end{align}
where $E = E_\mathrm{F}$ is chosen as the real Fermi energy of the Chern insulator ($E_\mathrm{F} = 0$ in our Dirac theory).
Notably, Eq.~(\ref{eq: perturbedChernSlabHam}) coincides with the continuum limit of the Hatano-Nelson model~\cite{Hatano1996, Hatano1997}, in which $b$ denotes the asymmetry of the hopping amplitudes. 
While the Pauli matrices $\tau_i$ in the Hatano-Nelson model describe the valley degrees of freedom, $\tau_i$ in our model describe the different flavors of chiral edge modes at the different boundaries.
We also note that this definition involves integration bounds at $k_x \rightarrow \pm \infty$ that are appropriate for the low-energy continuum theory in Eq.~\eqref{eq: perturbedChernSlabHam}. 
For a microscopic lattice model, we should instead integrate over the 1D Brillouin zone $k_x \in [0,2\pi)$ and include the contributions from the bulk modes. 
In such a case, the chiral edge modes connect with the bulk modes, making the complex-spectral winding number $W(E)$ well defined and consistent with the continuum description.

\begin{figure}[t]
\centering
\includegraphics[width=\linewidth]{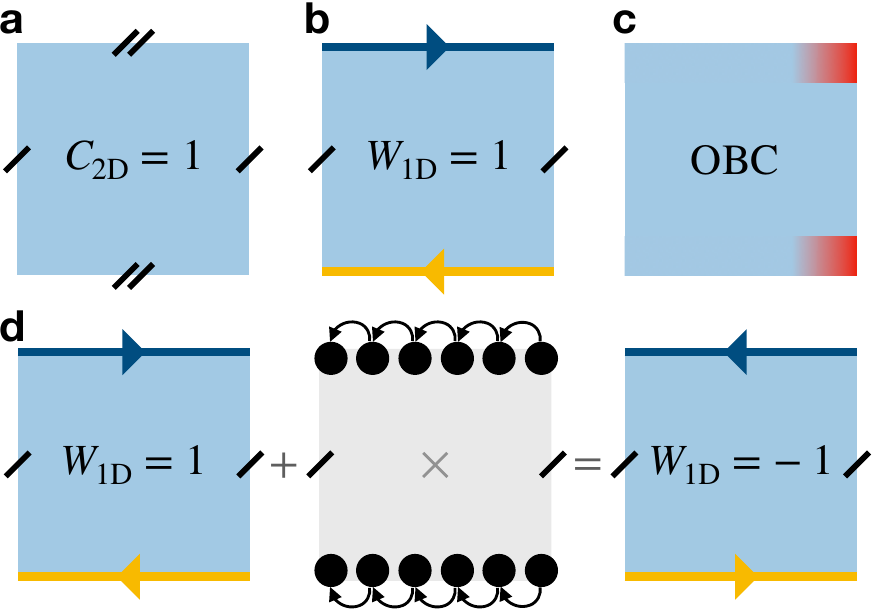}
\caption{Skin effect at the boundaries of a non-Hermitian Chern insulator. (a)~A Chern insulator with periodic boundary conditions is characterized by a 2D Hermitian topological invariant, the Chern number $C$. We here consider the case for $C = 1$. (b)~With open boundary conditions in the vertical direction (slab geometry), the Chern insulator hosts two counter-propagating edge modes, one for each edge. In the presence of small non-Hermitian perturbations and pseudo-inversion symmetry, these edge modes acquire opposite lifetimes (i.e., their energies develop opposite imaginary parts), here indicated in blue and yellow. Viewing the Chern insulator slab as an effective 1D system extending along the horizontal, this induces a nontrivial 1D non-Hermitian topological invariant, the complex-spectral winding number $W=1$. (c)~In open boundary conditions along both directions, $W=1$ induces a skin effect that manifests at two corners of the system, here indicated in red. (d)~Gluing a Hatano-Nelson chain (shown in black with unidirectional hopping indicated by the arrows) to each of the the two edges reduces the non-Hermitian winding number $W$ by $2$. Hence, $W=C$ holds only modulo $2$. This illustrates the importance of pseudo-inversion symmetry (with the inversion center indicated by the gray cross in the center panel): without pseudo-inversion symmetry, it would be possible to couple a single Hatano-Nelson chain to the boundary and thereby trivialize the non-Hermitian winding number $W$ and skin effect.}
\label{fig: chern_schematic}
\end{figure}

While the complex-spectral winding number $W(E)$ can be nonzero in general, we now show that additional spatial symmetry can guarantee $W(E) \neq 0$. In particular, we require pseudo-inversion symmetry
\begin{equation}
\mathcal{I} H(\bs{k}) \mathcal{I}^\dagger = H(-\bs{k})^\dagger,
    \label{eq: pseudo inversion}
\end{equation}
which exchanges the top and bottom edges. In the particular Dirac model in Eq.~\eqref{eq: perturbedChernSlabHam}, we have $\mathcal{I} = \tau_x$.
Inversion symmetry is fundamental for topological phases by yielding symmetry indicators~\cite{FuKane2007} and protecting topological crystalline phases~\cite{Hughes2011}.
Pseudo-inversion symmetry is a generalization of inversion symmetry in NH systems and similarly relevant to NH topological phases~\cite{Okugawa2021, Vecsei2021, ShiozakiOno2021}.
Pseudo-inversion symmetry is respected by the NH chiral edge modes as long as it is respected by the NH Chern insulator in the 2D bulk, and is compatible with other internal and spatial symmetries, as well as the nontrivial Chern number.
In the presence of pseudo-inversion symmetry, we must have $a = 0$ while $b$ can be arbitrary.
In this case, the winding number $W(0)$ is always nonzero and given by $W(0) = -\mathrm{sign}(v b)$, as long as there is a point gap at $E = 0$ (that is, as long as $b \neq 0$). While we cannot predict the sign of the winding number, we have thereby shown $|W(0)|=1$ as long as pseudo-inversion-symmetric NH perturbations are present. 
We also note that the complex-spectral winding number vanishes in the presence of conventional inversion symmetry $\mathcal{I} H(\bs{k}) \mathcal{I}^\dagger = H(-\bs{k})$~\cite{Kawabata2019, Vecsei2021}.

By the bulk-boundary correspondence of NH topological systems, the complex-spectral winding number $W(E)$ in Eq.~(\ref{eq: 1D winding}) leads to the skin effect under OBC~\cite{ZhangYangFang2020, Okuma2020_toposkin}.
Consequently, $W(E) \neq 0$ for the NH chiral edge modes results in second-order skin modes localized at the corners.
Importantly, this corner skin effect is guaranteed by the NH topology in the 2D bulk---the intricate combination between the line-gap topology (i.e., Chern number) and the NH perturbation that opens the point gap, supplemented with pseudo-inversion symmetry (see also Appendix~\ref{appendix: fragile}).
Hence, it is stable against, for example, symmetry-preserving continuous deformations at boundaries.
It should also be noted that the corner skin effect manifests itself in a semi-infinite system that is concerned solely with one boundary.
In other words, the corner skin effect can be captured by pseudospectra instead of spectra~\cite{Trefethen-Embree-textbook}, as is also the case for the conventional (i.e., first-order) skin effect~\cite{Okuma2020_toposkin}.
In a finite system with open boundaries, by contrast, the two NH chiral edge modes can be coupled with each other, and the skin modes may appear in arbitrary boundary segments, as shown with a lattice model shortly. Still, when the sample is cut into a rectangular geometry of linear extent $L$, $\mathcal{O}(L)$ chiral modes generically accumulate at corners rather than edges in the absence of fine tuning, analogous to 2D higher-order topological insulators that feature gapless 1D edges only for certain fine-tuned edge alignments~\cite{Schindler2020}.

Let us now turn to a system with $C = 2$. The unperturbed low-energy Hamiltonian is given by
\begin{equation} \label{eq: chern2edgetheory}
H_0(k_x) = v k_x \tau_z \sigma_0,
\end{equation}
where $\sigma_i$ ($i=0,x,y,z$) is another set of Pauli matrices that acts on the two chiral modes of each edge. Preserving pseudo-inversion symmetry $\mathcal{I} = \tau_x \sigma_0$, NH perturbations may for instance result in the NH Hamiltonian
\begin{equation}
H(k_x) = v k_x \tau_z \sigma_0 + \mathrm{i} a \tau_z \sigma_0 + \mathrm{i} b \tau_z \sigma_z,
\end{equation}
with $a,b \in \mathbb{R}$. 
For $a > 0$ and $b = 0$, we now obtain $W(0) = 2$, while for $a = 0$ and $b > 0$, we obtain $W(0) = 0$, realizing a counter-example to the naive conjecture that the boundary NH winding number may be fully determined by the bulk Chern number. 
This example is sufficient to establish that pseudo-inversion symmetry only fixes the slab geometry winding number modulo $2$, consistent with the topological classification~\cite{Kawabata2019}. 
The same observation may also be understood as follows [Fig.~\ref{fig: chern_schematic}\,(d)]. Without undergoing a bulk phase transition, we can glue a Hatano-Nelson chain with $W(0)=\pm 1$ to the top boundary of the sample ($y>0$). Pseudo-inversion symmetry then implies that we should also glue another Hatano-Nelson chain \emph{with the same winding number} to the bottom boundary ($y<0$). This adiabatic process changes the net winding number of the slab geometry by $\pm 2$. Correspondingly, $W(0)$ can be determined only modulo $2$ from a bulk topological invariant in the presence of pseudo-inversion symmetry.

\begin{figure}[t]
\centering
\includegraphics[width=\linewidth]{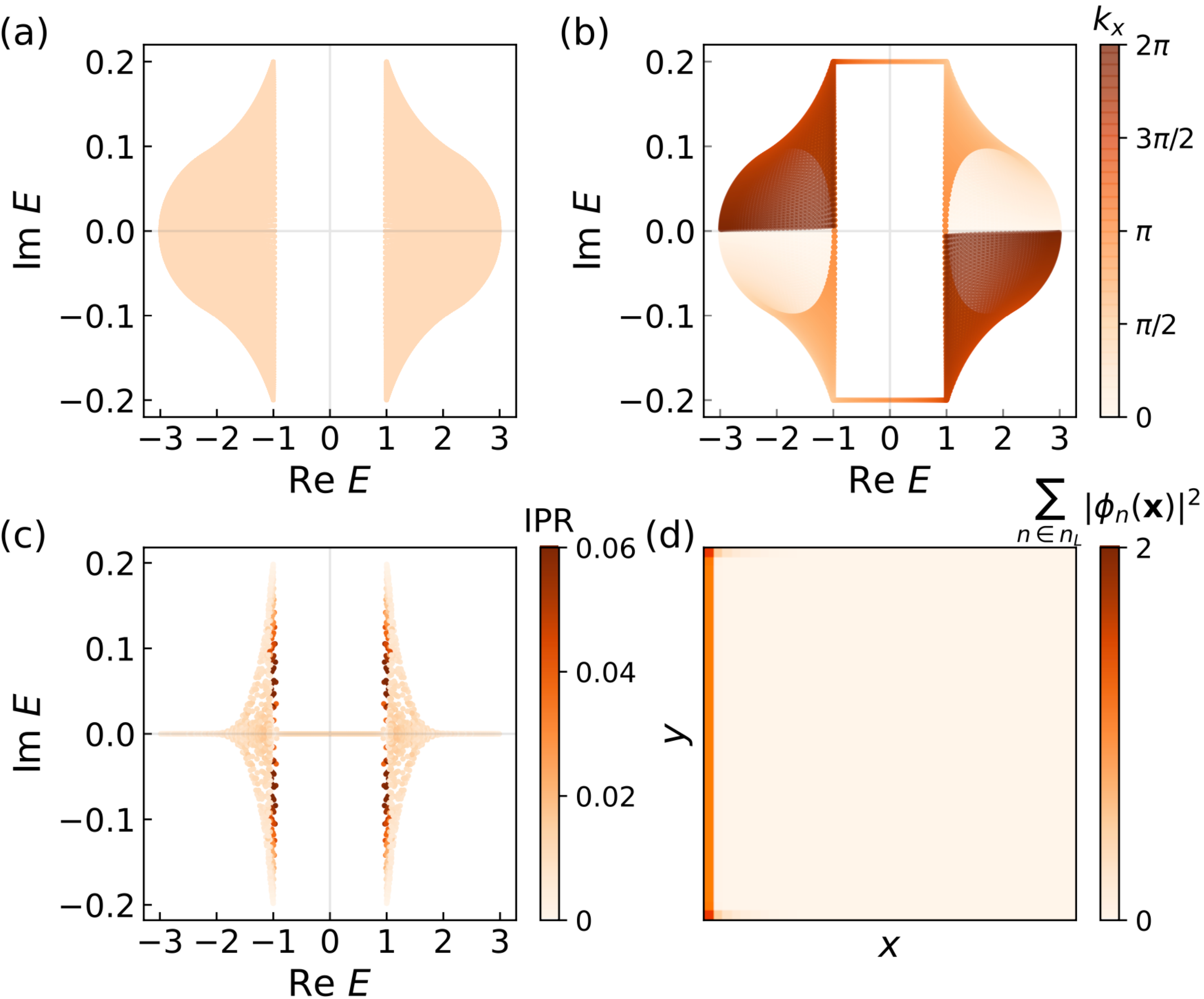} 
\caption{Non-Hermitian boundary topology in a Chern insulator [$L_x = L_y = 40$, $t=1.0$, $m = 1.0$, $\gamma = 0.2$ in Eq.~\eqref{eq: chern_lattice_model}].
(a)~Complex spectra for periodic boundary conditions in both $x$ and $y$-directions, (b)~slab geometry (periodic boundary conditions in the $x$-direction and open boundary conditions in the $y$-direction), and (c)~open boundary conditions in both $x$ and $y$-directions. Color bar in (c): inverse participation ratio (IPR) $\sum_{\bm{x}} \left| \phi_n \left( \bm{x} \right) \right|^4/\left( \sum_{\bm{x}} \left| \phi_n \left( \bm{x} \right) \right|^2 \right)^2$ for each eigenmode $\phi_n \left( \bm{x} \right)$. (d)~Average spatial profile of eigenmodes under open boundary conditions in both $x$ and $y$-directions, where $n_{L}$ denotes the set of in-gap states with energy $\mathrm{Re}\,E\in (-1,1)$.
}
\label{fig: Chern}
\end{figure}

As a consequence, the 1D NH winding number $W(E)$ of the slab geometry satisfies
\begin{equation} \label{eq: chern_winding_correspondence}
    W(E_\mathrm{F}) = C \quad (\mathrm{mod}\ 2)
\end{equation}
for a NH insulator in class A that is point-gapped at the Fermi level $E_\mathrm{F}$ and preserves pseudo-inversion symmetry (Tab.~\ref{tab: mapping2D}). Correspondingly, each edge of a Chern insulator realizes an anomalous \emph{half} of a Hatano-Nelson chain. Around the Fermi level $\mathrm{Re}\,E = E_{\rm F}$, the Hatano-Nelson chain gives rise to two valley degrees of freedom (i.e., right and left movers), whose imbalance in lifetime is caused by non-Hermiticity.
Notably, the NH chiral edge modes in Eq.~(\ref{eq: perturbedChernSlabHam}) have the same complex spectrum and topological features as the Hatano-Nelson chain in the long wavelength limit.
Specifically, in the slab geometry, each of the NH chiral edge modes, as well as each of the valley degrees of freedom in the Hatano-Nelson chain, gives rise to the complex-spectral winding number $\left| W(E_\mathrm{F}) \right| = 1/2$, the sum of which yields $\left| W(E_\mathrm{F}) \right| = 1/2 \times 2 = 1$.
An important difference is, however, that the NH chiral edge modes appear only at the boundaries, leading to a new type of NH boundary phenomena---a higher-order skin effect.

\subsection{Lattice model}

We confirm the above continuum description by a NH Chern insulator on a square lattice, described by~\cite{Shen2018, YaoSongWang2018_Chern, Kunst2018, Kawabata2018_Chern, Yokomizo2022}
\begin{align} \label{eq: chern_lattice_model}
    H \left( \bm{k} \right) &= \left( m+t\cos k_x + t\cos k_y \right) \sigma_x \nonumber \\ 
    &\qquad\quad+ \left( t \sin k_x + \ii \gamma \right) \sigma_y + \left( t\sin k_y \right) \sigma_z
\end{align}
with $t, m, \gamma \in \mathbb{R}$.
Under PBC, the complex spectrum is obtained as [Fig.~\ref{fig: Chern}\,(a)]
\begin{align}
    E \left( \bm{k} \right) &= \pm \left[ \left( m+t\cos k_x + t\cos k_y \right)^2 \right. \nonumber \\ 
    &\qquad\quad \left. + \left( \ii \gamma +t \sin k_x \right)^2 + \left( t\sin k_y \right)^2 \right]^{1/2}.
\end{align}
This NH lattice model indeed respects pseudo-inversion symmetry in Eq.~(\ref{eq: pseudo inversion}) with $\mathcal{I} = \sigma_x$.
Furthermore, it is also invariant under the combination of time reversal$^{\dag}$ and reflection along the $x$-direction:
\begin{align}
    \mathcal{M}_y H (k_x, k_y) \mathcal{M}_y^{\dag} = H (k_x, -k_y)^{\dag},
        \label{eq: 2D mirror-time}
\end{align}
where $\mathcal{M}_y = \sigma_x \mathit{K}$ is an antiunitary operator with complex conjugation $\mathit{K}$.
Notably, this symmetry forbids the skin effect along the $y$-direction in the bulk~\cite{Kawabata2019}.
On the other hand, no such constraints are present along the $x$-direction. We will later discuss the effect of relaxing this symmetry, which is not necessary for Eq.~\eqref{eq: chern_winding_correspondence} but convenient for our numerics since it allows us to clearly separate bulk and boundary skin effects.

In the absence of the NH perturbation (i.e., $\gamma = 0$), this model is characterized by the nontrivial Chern number $C = \mathrm{sgn} \left( m/t \right)$ for $\left| m/t \right| < 2$ and no Chern number $C = 0$ for $\left| m/t \right| > 2$.
Also in the presence of the NH perturbation, the topologically nontrivial phase persists as long as the real line gap is open.
Consequently, a pair of chiral edge modes appears under PBC along the $x$-direction and OBC along the $y$-direction [Fig.~\ref{fig: Chern}\,(b)], whose energy dispersion is exactly obtained as~\cite{Kawabata2018_Chern}
\begin{align}
    E \left( k_x \right) = \pm \left( t \sin k_x + \ii \gamma \right).
\end{align}
Here, the sign $\pm$ changes depending on the different boundaries, and the momentum cutoff $\left| \cos k_x + m/t \right| < 1$ is imposed.
These chiral edge modes on a lattice reduce to Eq.~(\ref{eq: perturbedChernSlabHam}) in the continuum limit $k_x \to 0$, where $t$ and $\gamma$ correspond to $v$ and $b$, respectively.
Consistent with the continuum description, one of the chiral edge modes has a positive imaginary part of eigenenergy and the other has a negative imaginary part, leading to the opening of a point gap around $E = 0$. The combined system of the bulk and chiral edge modes contributes to the complex-spectral winding number $\left| W \right| = 1$ in Eq.~(\ref{eq: 1D winding}).
We note that no skin effect occurs for the bulk modes under PBC along the $x$-direction and OBC along the $y$-direction due to the extra symmetry $\mathcal{M}_y$ in Eq.~(\ref{eq: 2D mirror-time}).

As a consequence of the complex-spectral winding, the skin effect occurs in full OBC (i.e., OBC along both $x$ and $y$-directions).
We obtain the complex spectrum for full OBC [Fig.~\ref{fig: Chern}\,(c)].
An extensive number of eigenstates in the bulk are subject to the skin effect, but localized only in the $x$-direction and delocalized in the $y$-direction.
This skin effect of bulk modes originates from the 1D weak point-gap topology~\cite{Okuma2020_toposkin, Hofmann2020}.
In fact, the bulk complex spectrum with fixed $k_y$ forms loops in the complex-energy plane, leading to the nontrivial complex-spectral winding as a function of $k_x$ for each fixed $k_y$.
By contrast, the 1D weak point-gap topology is trivial along the $y$-direction.
This is due to the spatial symmetry in Eq.~(\ref{eq: 2D mirror-time}), which forbids the skin effect along the $y$-direction in the bulk.
The different localization properties can be distinguished, for example, by the inverse participation ratios $\sum_{x, y} \left| \phi_n \left( x, y \right) \right|^4/\left( \sum_{x, y} \left| \phi_n \left( x, y \right) \right|^2 \right)^2$, where the bulk skin modes (corner skin modes) have $O \left( 1/L_y \right)$ [$O \left( 1 \right)$] inverse participation ratios.
Some special eigenmodes at the edges of the bulk spectrum are localized at the corners. Since they are intermediate eigenmodes between the bulk modes and chiral edge modes, they are not expected to exhibit general or universal behavior.

On the other hand, the NH chiral edge modes also exhibit the skin effect. While their spectrum is complex in the slab geometry, it becomes entirely real in full OBC [compare Fig.~\ref{fig: Chern}\,(c) with Fig.~\ref{fig: Chern}\,(b)].
All of these gapless modes that cross the real line gap are localized along the $x$-direction as a consequence of the nontrivial complex-spectral winding number in the slab geometry [Fig.~\ref{fig: Chern}\,(d)].
By contrast, they can be delocalized along the $y$-direction in our model.
As also discussed above, this is not a general feature but due to the symmetry in Eq.~\eqref{eq: 2D mirror-time} that is not necessary to establish the correspondence in Eq.~(\ref{eq: chern_winding_correspondence}). 
The NH chiral edge modes exhibit the corner skin effect if further generic perturbation terms that break this symmetry are added to the Hamiltonian.
In fact, if a term $+\mathrm{i} \gamma_2 \sigma_z$ is added to Eq.~\eqref{eq: chern_lattice_model}, which respects pseudo-inversion symmetry and is the only constant NH perturbation other than the existing term multiplying $\gamma$, the reflection-type symmetry in Eq.~(\ref{eq: 2D mirror-time}) is broken. 
In this case, a generic NH skin effect appears when open boundaries are introduced in the $y$-direction. 
Importantly, however, the point gap of the slab spectrum remains open and its nontrivial winding number $\left| W \right| = 1$ remains quantized. 
Correspondingly, when OBC are also introduced in the $x$-direction, a topologically-protected higher-order NH skin effect occurs for which almost all bulk modes are localized at a corner. 
Notably, such an anomalous localization only at a single corner is compatible with pseudo-inversion symmetry, because this symmetry maps not between right eigenvectors of the NH Hamiltonian but rather maps between right and left eigenvectors.

The skin effects of bulk modes and chiral edge modes arise from different topological origins and hence can be compatible with each other in a single system.
While the former originates from the 1D weak point-gap topology, the latter originates from our correspondence between the Hermitian bulk and NH boundary.
Below, we show that this correspondence is generally applicable and yields several different types of NH boundary phenomena.
Owing to the different topological origins, the two types of skin effect can be clearly separated.
One convenient way is to add the NH perturbations only at boundaries, while we have hitherto focused on homogeneous NH perturbations throughout the system.
In such a case, a point gap can be open only at the boundaries, and hence only the corner skin effect arises.
This procedure generally applies to the subsequent examples in this work, for arbitrary symmetry classes and arbitrary spatial dimensions.

\section{2D \texorpdfstring{$\mathbb{Z}_2$}{Z2}-classified topological insulator (class AII)} 
    \label{sec: 2DTI_story}
    
 While time-reversal symmetry renders the Chern number trivial $C=0$, it can instead give rise to a $\mathbb{Z}_2$-valued topological invariant $\nu \in \{0,1\}$~\cite{Kane-Mele-05-Z2, Fu-Kane-06, Moore-Balents-07, Hasan-Kane-review, Qi-Zhang-review, Schnyder-Ryu-review}.
This $\mathbb{Z}_2$ topological invariant, protected by time-reversal symmetry, physically induces the quantum spin Hall effect accompanied by the emergence of helical edge modes.
Here, we study 2D time-reversal-invariant Hermitian systems in class AII and the corresponding NH boundary phenomena.

At low energy, the Hermitian slab Hamiltonian hosts a pair of helical edge modes at each boundary described by
\begin{equation}
    H_0(k_x) = v k_x \tau_z \sigma_x.
        \label{eq: 1D helical Hermitian}
\end{equation}
Here, the Pauli matrix $\sigma_i$ describes the spin degree of freedom, and time-reversal symmetry is represented by 
\begin{equation}
  \mathcal{T} H_0(\bs{k}) \mathcal{T}^\dagger = H_0(-\bs{k})
    \label{eq: TRS}
\end{equation}
with $\mathcal{T} = \mathrm{i} \tau_0 \sigma_y \mathit{K}$ and complex conjugation $K$. 
Similarly to the previous case, $\tau_z = +1$ ($\tau_z = -1$) describes a pair of the helical edge modes at the top (bottom) boundary.
We again set the Fermi energy at $E_\mathrm{F} = 0$.

Time-reversal symmetry in Eq.~(\ref{eq: TRS}) is generalized to NH systems in two manners~\cite{Kawabata2019}.
One is time-reversal symmetry in Eq.~(\ref{eq: TRS}), which corresponds to class AII in NH systems.
The other is time-reversal symmetry$^{\dag}$,
\begin{equation}
  \mathcal{T} H(\bs{k}) \mathcal{T}^\dagger = H(-\bs{k})^\dagger,
    \label{eq: TRS-dag}
\end{equation}
which corresponds to class AII$^{\dag}$ in NH systems.
While Eqs.~(\ref{eq: TRS}) and (\ref{eq: TRS-dag}) are equivalent to each other for Hermitian Hamiltonians, this is not the case for NH Hamiltonians.
Notably, different topological invariants apply to the two different symmetry classes.
Consequently, the helical edge modes induce different types of NH skin effect in classes AII and AII$^{\dag}$, as shown below.

\subsection{NH class AII}
NH Hamiltonians in class AII are defined to respect time-reversal symmetry in Eq.~(\ref{eq: TRS}). The general NH perturbation in this symmetry class is 
\begin{equation}
H(k_x) = v k_x \tau_z \sigma_x + \mathrm{i} \sum_{i=x,y,z} (a_i \tau_0 + b_i \tau_z) \sigma_i,
\end{equation}
with $a_i, b_i \in \mathbb{R}$. This Hamiltonian has a spectrum
\begin{align}
    &E_{\mu \nu}(k_x) = \mu \left[ (v k_x + \mathrm{i} \nu a_x + \mathrm{i} b_x)^2 \right. \nonumber \\
    &\qquad\qquad\qquad\quad \left. - (\nu a_y + b_y)^2 - (\nu a_z + b_z)^2 \right]^{1/2},
\end{align}
where $\mu = \pm 1$ and $\nu = \pm 1$ cycle through the four different eigenvalues. 
For $a_x \neq 0$ and $b_x = 0$, the point gap at $E=0$ is open, but the complex-spectral winding number in Eq.~(\ref{eq: 1D winding}) vanishes, $W(0) = 0$. 
We here note that the complex-spectral winding number $W(0)$ can be trivial even in the presence of a point gap~\cite{ZhangYangFang2020, Okuma2020_toposkin}, as is also the case for our Dirac Hamiltonian with $a_x \neq 0$ and $b_x = 0$.
On the other hand, for $a_x = 0$ and $b_x \neq 0$, we find $W(0) = -2 \, \mathrm{sign} (v b_x)$.
Correspondingly, imposing pseudo-inversion symmetry with $\mathcal{I} = \tau_x \sigma_0$ leads to $a_i = 0$ for all $i = x, y, z$ and protects the nontrivial spectral winding. 
The result $|W(0)| = 2$ is compatible with the $2 \mathbb{Z}$ classification of 1D NH systems in class AII, where the odd winding numbers are forbidden~\cite{Kawabata2019}. 

As a consequence, the 1D complex-spectral winding number $W(E)$ of the helical edge states in the slab geometry satisfies
\begin{equation}
    W(E_\mathrm{F}) = 2 \nu \quad (\mathrm{mod}~4)
\end{equation}
for a NH system in class AII that is point-gapped at the Fermi level $E_\mathrm{F}$ and preserves pseudo-inversion symmetry (Tab.~\ref{tab: mapping2D}).

\subsection{NH class \texorpdfstring{AII$^\dagger$}{AII†}}
NH Hamiltonians in class AII$^{\dag}$ are defined to respect time-reversal symmetry$^{\dag}$ in Eq.~(\ref{eq: TRS-dag}).
The general NH perturbations in class AII$^{\dag}$ now result in the low-energy slab Hamiltonian
\begin{equation}
H(k_x) = v k_x \tau_z \sigma_x + \mathrm{i} a \tau_0 \sigma_0 + \mathrm{i} b \tau_z \sigma_0,
    \label{eq: Dirac 1D AIIdag}
\end{equation}
with $a,b \in \mathbb{R}$.
Because of time-reversal symmetry$^{\dag}$ in Eq.~(\ref{eq: TRS-dag}), the complex-spectral winding number in Eq.~(\ref{eq: 1D winding}) always vanishes, which is compatible with the $\mathbb{Z}_2$ topological classification in terms of point gaps~\cite{Kawabata2019}.
In fact, we have $W (E_{\rm F}) = 0$ for Eq.~(\ref{eq: Dirac 1D AIIdag}),
and instead have the nontrivial $\mathbb{Z}_2$ topological invariant
\begin{align}
    &\left( -1 \right)^{\nu \left( E \right)} = \mathrm{sgn} \left\{ \frac{\mathrm{Pf} \left[ H \left( k_x=\pi \right) \mathcal{T} \right]}{\mathrm{Pf} \left[ H \left( k_x=0 \right) \mathcal{T} \right]} \right. \nonumber \\
    &\qquad\quad \left. \times \exp \left[ - \frac{1}{2} \int^{k_x=\pi}_{k_x=0} d\log \det \left[ H \left( k_x \right) \mathcal{T} \right] \right] \right\}.
        \label{eq: 1D Z2}
\end{align}
Depending on the values of $a$ and $b$, the $\mathbb{Z}_2$ topological invariant may be trivial or nontrivial. 
This ambiguity is again resolved by imposing pseudo-inversion symmetry $\mathcal{I} = \tau_x \sigma_0$, which sets $a = 0$. We then find $\nu(0) = 1$ as long as the point gap at $E = 0$ is open (that is, as long as $b \neq 0$).

As a consequence, the 1D $\mathbb{Z}_2$ topological invariant $\nu(E)$ of the helical edge states in the slab geometry satisfies
\begin{equation}
    \nu(E_\mathrm{F}) = \nu
\end{equation}
for a NH system in class AII$^\dagger$ that is point-gapped at the Fermi level $E_\mathrm{F}$ and preserves pseudo-inversion symmetry (Tab.~\ref{tab: mapping2D}).
This relationship is guaranteed also by K-theory~\cite{Kawabata2019}.

Notably, as a consequence of the different topological invariants, different types of corner skin effect appear.
In class AII, the NH helical edge states are characterized by the complex-spectral winding number in Eq.~(\ref{eq: 1D winding}), similarly to the chiral edge states in a NH Chern insulator in Sec.~\ref{sec: Chern_story}.
Hence, the skin modes are localized at the same side of the sample in full OBC.
In class AII$^{\dag}$, on the other hand, the NH helical edge states are no longer characterized by the winding number in Eq.~(\ref{eq: 1D winding}) but exhibit the $\mathbb{Z}_2$ topological invariant in Eq.~(\ref{eq: 1D Z2}).
An important feature of the skin effect induced by this $\mathbb{Z}_2$ topological invariant is the presence of skin modes at both boundaries, depending on the spin degree of freedom~\cite{Okuma2020_toposkin}.
Consequently, in contrast to class AII, the skin effect occurs at all the corners in full OBC.
Below, we confirm these different types of corner skin effect in a NH lattice model.

\subsection{Lattice model}

\begin{figure}[t]
\centering
\includegraphics[width=\linewidth]{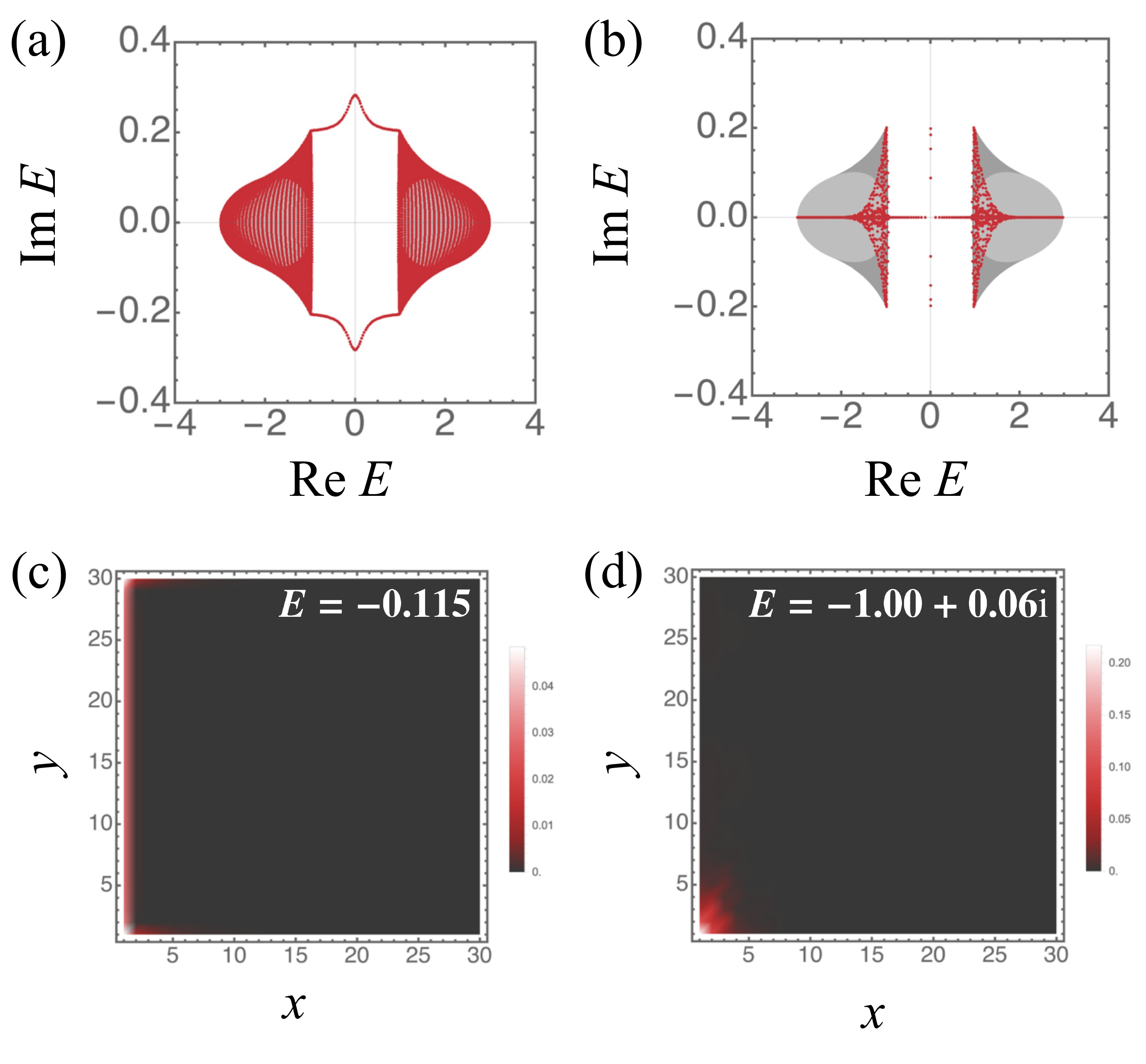} 
\caption{Non-Hermitian Bernevig-Hughes-Zhang model in class AII [$t=1.0$, $m=1.0$, $\gamma_{y0} = 0.2$, $\gamma_{zy} = 0.2$ in Eqs.~\eqref{eq: bhz_basic} and  \eqref{eq: aii_perturbations}].
(a, b)~Complex spectra under the periodic boundary conditions along both $x$ and $y$-directions (gray regions), slab geometry (periodic boundary conditions along the $x$-direction and open boundary conditions along the $y$-direction) [red dots in (a)], and open boundary conditions along both $x$ and $y$-directions [red dots in (b)].
(c, d)~Spatial profiles of eigenstates for (c)~$E=-0.115$ and (d)~$E=-1.00+0.06\ii$.
The system sizes are chosen as (a)~$200 \times 30$ sites and (b, c, d)~$30 \times 30$ sites.
}
	\label{fig: NH-2D-AII}
\end{figure}

As a lattice model for the above continuum description, we study the Bernevig-Hughes-Zhang (BHZ) model~\cite{BHZ-06}
\begin{align} \label{eq: bhz_basic}
    H_0 \left( \bm{k} \right) &= \left( m+t\cos k_x + t\cos k_y \right) \tau_x \sigma_0 \nonumber \\ 
    &\qquad\quad + \left( t \sin k_x \right) \tau_y \sigma_0 + \left( t\sin k_y \right) \tau_z \sigma_x
\end{align}
with $t, m \in \mathbb{R}$.
The BHZ model respects time-reversal symmetry in Eq.~(\ref{eq: TRS}) with $\mathcal{T} = \ii \tau_0 \sigma_y K$ and exhibits the $\mathbb{Z}_2$-nontrivial ($\mathbb{Z}_2$-trivial) topological invariant $\nu = 1$ ($\nu = 0$) for $\left| m/t \right| < 2$ ($\left| m/t \right| > 2$).
Consequently, under PBC along the $x$-direction and OBC along the $y$-direction, a pair of helical edge states appears at each boundary, which reduce to the energy dispersion in Eq.~(\ref{eq: 1D helical Hermitian}) in the low-energy limit $k_x \to 0$.
The BHZ model also respects inversion symmetry in Eq.~(\ref{eq: pseudo inversion}) with $\mathcal{I} = \tau_x\sigma_0$.

\begin{figure}[t]
\centering
\includegraphics[width=\linewidth]{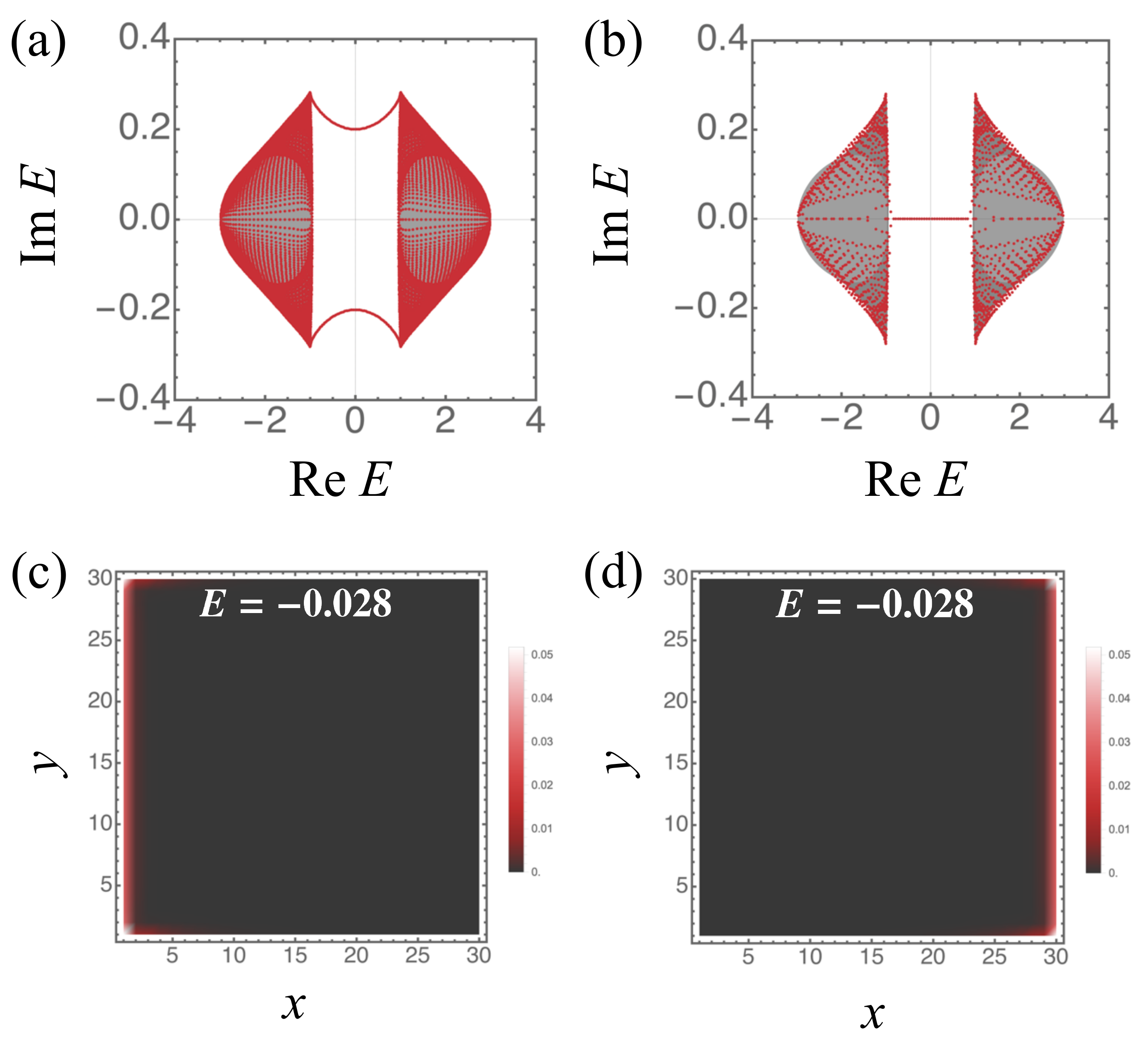} 
\caption{Non-Hermitian Bernevig-Hughes-Zhang model in class AII$^{\dag}$ [$t=1.0$, $m=1.0$, $\gamma_{yx} = 0.2$, $\gamma_{yz} = 0.2$ in Eqs.~\eqref{eq: bhz_basic} and~\eqref{eq: aiidagger_perturbations}].
(a, b)~Complex spectra under the periodic boundary conditions along both $x$ and $y$-directions (gray regions), slab geometry (periodic boundary conditions along the $x$-direction and open boundary conditions along the $y$-direction) [red dots in (a)], and open boundary conditions along both $x$ and $y$-directions [red dots in (b)].
(c, d)~Spatial profiles of Kramers-degenerate eigenstates for $E=-0.028$.
The system sizes are chosen as (a)~$200 \times 30$ sites and (b, c, d)~$30 \times 30$ sites.
}
	\label{fig: NH-2D-AIIdag}
\end{figure}

Now, we add NH perturbations to the BHZ model that preserve certain symmetries.
First, we consider the NH perturbations
\begin{align} \label{eq: aii_perturbations}
    \Delta H_{\rm AII} = \ii \gamma_{y0} \tau_y \sigma_0 + \ii \gamma_{zy} \tau_z \sigma_y
\end{align}
with $\gamma_{y0}, \gamma_{zy} \in \mathbb{R}$.
The combined NH system $H_0 \left( \bm{k} \right) + \Delta H_{\rm AII}$ respects time-reversal symmetry in Eq.~(\ref{eq: TRS}) with $\mathcal{T} = \ii \tau_0 \sigma_y K$ and hence belongs to class AII.
It also respects pseudo-inversion symmetry in Eq.~(\ref{eq: pseudo inversion}) with $\mathcal{I} = \tau_x\sigma_0$, as well as the spatial symmetry in Eq.~(\ref{eq: 2D mirror-time}) with $\mathcal{M}_y = \tau_x K$.
Under OBC only along the $y$-direction, a pair of helical edge modes appears, and a point gap is open, further leading to the skin effect under full OBC (Fig.~\ref{fig: NH-2D-AII}).
Similarly to the NH Chern insulator in 2D class A, the NH helical edge modes are generically localized at an edge, instead of the corners. As also discussed above, this is due to the additional spatial symmetry; under further generic perturbations, the NH helical edge modes are localized at the corner and exhibit the second-order skin effect.

Next, we consider the NH perturbations that preserve different symmetry, 
\begin{align} \label{eq: aiidagger_perturbations}
    \Delta H_{\rm AII^{\dag}} = \ii \gamma_{yx} \tau_y \sigma_x + \ii \gamma_{yz} \tau_y \sigma_z
\end{align}
with $\gamma_{yx}, \gamma_{yz} \in \mathbb{R}$.
Instead of time-reversal symmetry in Eq.~(\ref{eq: TRS}), this NH generalization of the BHZ model, $H_0 \left( \bm{k} \right) + \Delta H_{\rm AII^{\dag}}$, respects time-reversal symmetry$^{\dag}$ with $\mathcal{T} = \ii \tau_0 \sigma_y K$ and hence belongs to class AII$^{\dag}$.
It also respects pseudo-inversion symmetry in Eq.~(\ref{eq: pseudo inversion}) and the spatial symmetry in Eq.~(\ref{eq: 2D mirror-time}). Similarly to the previous cases, a point gap is open under OBC only along the $y$-direction, and the skin effect occurs in full OBC (Fig.~\ref{fig: NH-2D-AIIdag}).
A unique feature of this symmetry class and the $\mathbb{Z}_2$ topological invariant manifests itself in the locations at which the skin modes are localized.
In fact, depending on the spin degree of freedom, the skin modes appear at both edges.
In the semi-infinite boundary conditions, a reciprocal pair of second-order skin modes appears at different corners.

\section{3D \texorpdfstring{$\mathbb{Z}$}{Z}-classified topological insulator (class AIII)}
    \label{sec: 3DTI_story}

Similarly to the 2D case, boundary modes of a 3D topological insulator or superconductor generally exhibit point-gap topology in the presence of NH perturbations.
In general, 3D Hermitian systems can host $\mathbb{Z}$ topological phases in AZ symmetry classes AIII, DIII, and CI, and $\mathbb{Z}_2$ topological phases in symmetry classes AII and CII (see Tab.~\ref{tab: mapping3D}).
As a prototypical example of the $\mathbb{Z}$ topological phase, we here study the Dirac surface modes of a 3D topological insulator protected by chiral symmetry in class AIII (Fig.~\ref{fig: AIII_schematic}).
In contrast to the 2D case, the NH boundary modes do not exhibit the skin effect but form chiral hinge modes.
These chiral hinge modes are stabilized by a point gap, rather than the line gap of second-order Hermitian topological insulators, and are hence unique to NH systems. In Sec.~\ref{sec: 3DZ2}, we separately explore non-Hermiticity in $\mathbb{Z}_2$-classified topological phases in 3D for symmetry class AII.

\subsection{Continuum theory}

\begin{figure}[t]
\centering
\includegraphics[width=\linewidth]{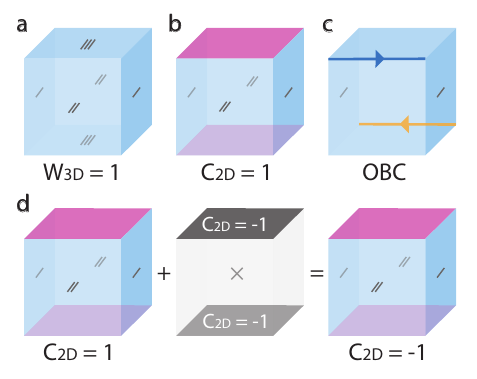}
\caption{Chiral hinge modes of a 3D non-Hermitian topological insulator with chiral symmetry (class AIII). (a)~In periodic boundary conditions, a 3D topological insulator protected by chiral symmetry is characterized by the Hermitian 3D winding number $W$. We here consider the case for $W = 1$. (b)~With open boundary conditions along the $z$-direction, the 3D topological insulator hosts two Dirac surface modes, one for each surface (indicated in purple). 
In the presence of non-Hermitian perturbations, the two surface modes together possess a point gap and exhibit the nontrivial 2D non-Hermitian topological invariant, $C=1$~\cite{Kawabata2019}. (c)~In open boundary conditions along both $z$ and $y$-directions, $C = 1$ leads to the emergence of chiral hinge modes, here indicated in blue and yellow. (d)~Gluing a 2D point-gapped non-Hermitian model (shown in black) to each of the two surfaces reduces the non-Hermitian topological invariant $C$ by $2$. Hence, $C=W$ holds only modulo $2$. This illustrates the importance of pseudo-inversion symmetry (with the inversion center indicated by the gray cross in the center panel): without pseudo-inversion symmetry, it would be possible to couple a single 2D point-gapped non-Hermitian model to one of the surfaces and thereby trivialize $C$ and the chiral hinge modes.}
\label{fig: AIII_schematic}
\end{figure}

The topological invariant of 3D Hermitian systems in class AIII is the 3D winding number $W \in \mathbb{Z}$~\cite{Schnyder-Ryu-08, Hasan-Kane-review, Qi-Zhang-review, Schnyder-Ryu-review}.
Here, Hermitian Hamiltonians in class AIII are defined to respect chiral symmetry
\begin{align}
    \CS H_0(\bs{k}) \CS^\dagger = - H_0(\bs{k}),
        \label{eq: CS}
\end{align}
with a unitary matrix $\CS$. For the simplest nontrivial case $W=1$, the low-energy continuum Hamiltonian in the slab geometry [Fig.~\ref{fig: AIII_schematic}\,(b)] is given by the Dirac surface modes
\begin{align}
    H_0(\bs{k}) = v k_x \tau_z \sigma_x + v k_y \tau_0 \sigma_y, 
        \label{eq: 2D chiral Dirac}
\end{align}
where the Pauli matrices $\sigma_i$ describe the internal degree of freedom at each surface, and $\tau_z = +1$ ($-1$) corresponds to the top (bottom) surface.
These surface modes indeed respect chiral symmetry in Eq.~(\ref{eq: CS}) with $\CS = \tau_0 \sigma_z$.

\begin{figure}[t]
\centering
\includegraphics[width=\linewidth]{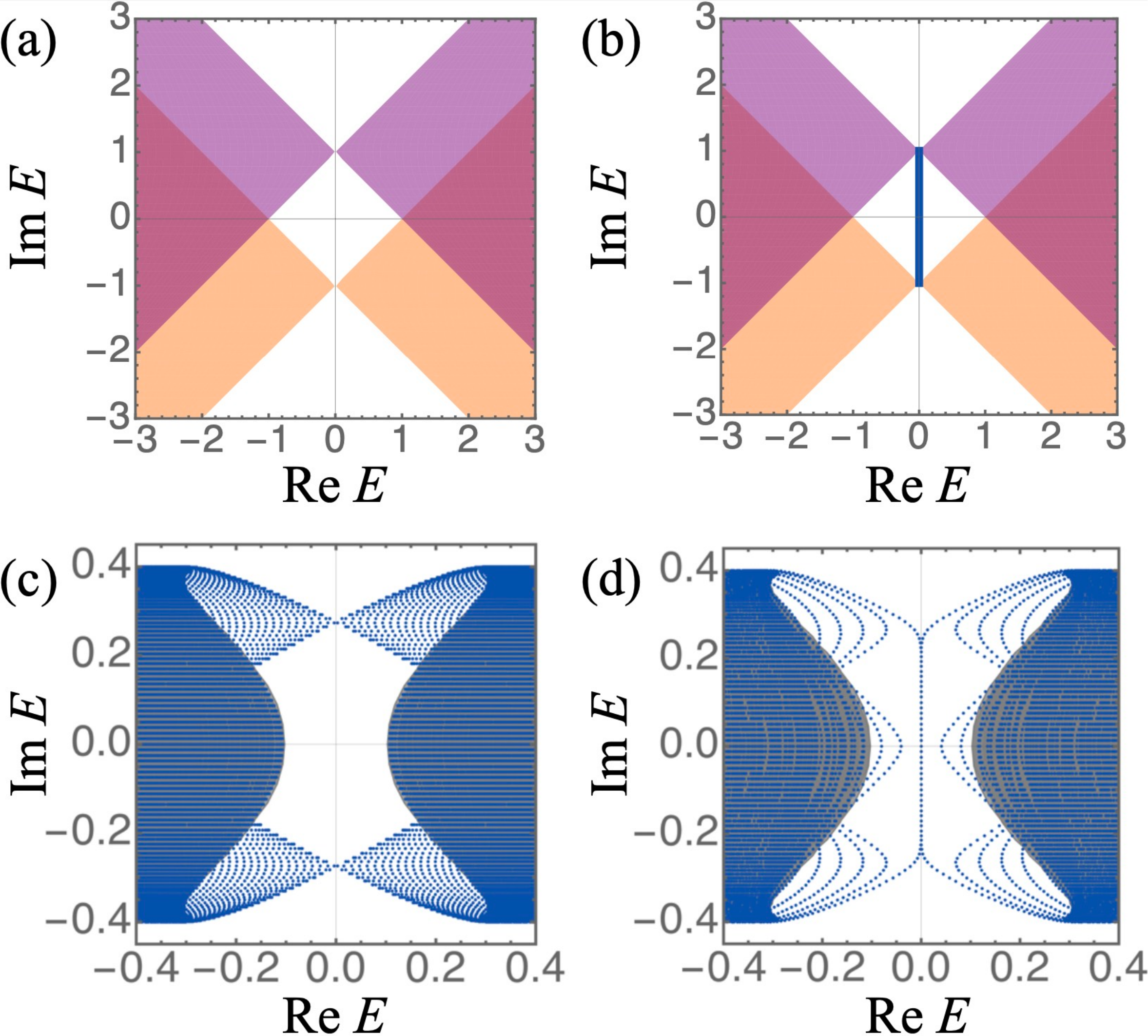} 
\caption{Non-Hermitian boundary topology in a 3D topological insulator with chiral symmetry (class AIII).
(a, b)~Complex spectra of non-Hermitian Dirac surface modes under the open boundary conditions along the (a)~$z$-direction and (b)~$y$ and $z$-directions [$v=v'=\delta=1.0$ in Eq.~\eqref{eq: 3D NH Ham AIII}].
The orange and purple regions denote the complex spectra of the Dirac surface modes at the different boundaries along the $z$-direction, and the blue line denotes the spectrum of the chiral hinge modes $E \left( k_x \right) = \ii v' k_x$.
(c, d)~Complex spectra of the non-Hermitian lattice model in 3D under the periodic boundary conditions (gray regions) and the open boundary conditions (blue dots) along the (c)~$z$-direction and (d)~$y$ and $z$-directions [$t_1=1.0$, $t_2=0.5$, $m=2.0$, $\delta=0.2$, $\gamma=0.4$ in Eq.~\eqref{eq: aiii_lattice_model}].
The system sizes are chosen as (c)~$200 \times 200 \times 30$ sites and (d)~$200 \times 30 \times 30$ sites.
}
\label{fig: NH-3D-AIII}
\end{figure}

We now consider NH perturbations in class AIII, which respect~\cite{Kawabata2019}
\begin{align}  
    \CS H(\bs{k}) \CS^\dagger = - H(\bs{k})^{\dag}.
        \label{eq: CS-NH}
\end{align}
As in the 2D phases considered before, we again impose pseudo-inversion symmetry 
in Eq.~(\ref{eq: pseudo inversion}) with $\mathcal{I} = \tau_y \sigma_x$.
As an example of NH perturbations that preserve both chiral and pseudo-inversion symmetries, we study
\begin{align}
    H(\bs{k}) &= v k_x \tau_z \sigma_x + v k_y \tau_0 \sigma_y +\delta \tau_0 \sigma_x + \ii v' k_x \tau_0 \sigma_0
        \label{eq: 3D NH Ham AIII}
\end{align}
with $\delta, v' \in \mathbb{R}$.
Here, the Hermitian perturbation $\delta \tau_0 \sigma_x$ shifts the Dirac point at each surface, and the NH perturbation $\ii v' k_x \tau_0 \sigma_0$ results in an imaginary-valued energy dispersion.
As a combination of these perturbations, a point gap opens around $E = 0$.
In fact, the complex spectrum is obtained as 
\begin{align}
    E(\bs{k}) = \pm v \sqrt{\left( k_x\tau + \delta/v \right)^2 + k_y^2} + \ii v' k_x,
\end{align}
where $\tau = +1$ ($\tau = -1$) corresponds to the top (bottom) surface [Fig.~\ref{fig: NH-3D-AIII}\,(a)].
Importantly, the NH Dirac surface modes exhibit the nontrivial topology in terms of the point gap around $E=0$.
In general, 2D point-gapped NH Hamiltonians $H(\bs{k})$ with chiral symmetry are characterized by the Chern number of the Hermitian matrix $\ii H(\bs{k}) \CS$~\cite{Kawabata2019}.
For our NH Dirac surface Hamiltonian, we have
\begin{align}
    \ii H(\bs{k}) \CS = v k_x \tau_z \sigma_y - v k_y \tau_0 \sigma_x + \delta \tau_0 \sigma_y - v' k_x \tau_0 \sigma_z,
\end{align}
which reduces to a continuum limit of a Hermitian Chern insulator and hence yields $C \left( 0 \right) = -\mathrm{sgn} \left( \delta v v' \right)$. Moreover, the only local and $\bs{k}$-independent NH perturbations to Eq.~\eqref{eq: 2D chiral Dirac} allowed by symmetry read $+\mathrm{i} m_1 \tau_0 \sigma_z$, which does not open a gap in $\ii H(\bs{k}) \CS$, and $+\mathrm{i} m_2 \tau_z \sigma_0$, which again results in $|C|=1$. Correspondingly, we find the general relation
\begin{equation}
    C(0) = W \quad (\mathrm{mod}~2)
\end{equation}
between the 3D Hermitian topological invariant $W$ in class AIII and the 2D NH topological invariant $C(0)$ in class AIII, evaluated at the reference energy $E=0$ within the point gap as fixed by chiral symmetry (Tab.~\ref{tab: mapping3D}).

In contrast to the complex-spectral winding number in Eq.~(\ref{eq: 1D winding}) and its $\mathbb{Z}_2$ counterpart in Eq.~(\ref{eq: 1D Z2}), this point-gap topological invariant has an analog in the line-gap topology and does not lead to the skin effect.
In fact, under the open boundary conditions along the $y$-direction, the boundaries host the chiral hinge states with the imaginary-valued dispersion
\begin{align}
    E (k_x) = \ii v' k_x,
        \label{eq: 3D AIII chiral hinge}
\end{align}
which are protected by the point gap and chiral symmetry [Figs.~\ref{fig: AIII_schematic}\,(c) and \ref{fig: NH-3D-AIII}\,(b)].
We note that while chiral symmetry can also be generalized to NH Hamiltonians by Eq.~(\ref{eq: CS}), which corresponds to class AIII$^{\dag}$, Eq.~(\ref{eq: CS}) cannot give rise to any point-gap topology in 2D~\cite{Kawabata2019}.

\subsection{Lattice model}
We also confirm the above continuum theory by the NH lattice model in 3D,
\begin{align} \label{eq: aiii_lattice_model}
    &H(\bs{k}) = \left[ m + t_1 \left( \cos k_x + \cos k_y + \cos k_z \right) \right] \tau_y \sigma_0 \nonumber \\
    &\quad+ t_2 \tau_x \left[ \left( \sin k_x \right) \sigma_x + \left( \sin k_y \right) \sigma_y + \left( \sin k_z \right) \sigma_z \right] \nonumber \\
    &\quad+ \delta \left( \cos k_x + \cos k_y \right) \tau_y \sigma_y + \ii \gamma \left( \sin k_x \right) \tau_0 \sigma_0
\end{align}
with $t_1$, $t_2$, $m$, $\delta$, $\gamma \in \mathbb{R}$.
This NH lattice model respects chiral symmetry in Eq.~(\ref{eq: CS-NH}) with $\CS = \tau_z\sigma_0$ and pseudo-inversion symmetry in Eq.~(\ref{eq: pseudo inversion}) with $\mathcal{I} = \tau_y \sigma_0$.
The complex spectrum for PBC is obtained as
\begin{align}
    &E(\bs{k}) = \\
    &\pm \sqrt{\left[\Delta (\bs{k}) \pm \delta \left( \cos k_x + \cos k_y \right)\right]^2 + t_2^2 \sin^2 k_y} +\ii \gamma \sin k_x, \nonumber
\end{align}
with $\Delta (\bs{k}) \coloneqq [ ( m + t_1 ( \cos k_x + \cos k_y + \cos k_z ) )^2 + t_2^2 ( \sin^2 k_x + \sin^2 k_z) ]^{1/2}$.
The real line gap at $\mathrm{Re}\,E = 0$ is open for $\delta = \gamma = 0$ and remains open for small $\delta$ and $\gamma$.
For $\delta = \gamma = 0$, the 3D winding number is obtained as $\left| W \right| = 2$ for $\left| m \right| < \left| t_1 \right|$, $\left| W \right| = 1$ for $\left| t_1 \right| < \left| m \right| < 3\left| t_1 \right|$, and $\left| W \right| = 0$ for $3\left| t_1 \right| < \left| m \right|$~\cite{Schnyder-Ryu-08}.
Consequently, for $\left| t_1 \right| < \left| m \right| < 3\left| t_1 \right|$ ($\left| m \right| < \left| t_1 \right|$), one flavor (two flavors) of Dirac surface modes appears under OBC, which reduce to Eq.~(\ref{eq: 2D chiral Dirac}) around $\mathrm{Re}\,E = 0$.

We obtain the complex spectrum under OBC along the $z$-direction [Fig.~\ref{fig: NH-3D-AIII}\,(c)].
We note that no skin effect occurs in this model.
As a consequence of the nonzero 3D winding number in the bulk, the Dirac surface states appear at each boundary along the $z$-direction.
Here, an extensive number of eigenstates are delocalized through the bulk, and the Dirac surface modes are generally delocalized in the $xy$ plane but localized at the boundaries along the $z$-direction.
Owing to the NH perturbations for $\delta \neq 0$ and $\gamma \neq 0$, the line-gapless points of the two Dirac surface modes at the different boundaries are shifted, and a point gap is open around $E = 0$, consistent with the continuum theory.
Furthermore, these NH Dirac surface modes exhibit the nontrivial point-gap topology, i.e., a nonzero Chern number $|C(0)| = 1$ of $\ii H(\bs{k}) \CS$.
Consequently, chiral hinge modes appear under OBC along both $y$ and $z$-directions [Fig.~\ref{fig: NH-3D-AIII}\,(d)].
The chiral hinge modes due to the NH point-gap topology are delocalized only along the $x$-direction and localized in the $yz$ plane.

\subsection{Dimensional reduction} \label{sec: dimensional_reduction_explainer}

The imaginary-valued dispersion of chiral hinge modes in Eq.~(\ref{eq: 3D AIII chiral hinge}) is closely related to the anomalous dispersion of an isolated exceptional point~\cite{TerrierKunst_2020, eti, Kawabata2021, Nakamura2022} via dimensional reduction.
To see this, let us consider a four-dimensional (4D) Chern insulator~\cite{ZhangHu01, Qi2008} with non-Hermiticity in a similar manner to the previous cases for 2D and 3D topological phases.
Under OBC along one direction, each boundary hosts a single Weyl mode
\begin{align}
    H_0 (\bs{k}) = v \left( k_x \tau_z \sigma_x + k_y \tau_0 \sigma_y + k_z \tau_0 \sigma_z \right),
\end{align}
where $\tau_z = +1$ ($\tau_z = -1$) denotes the top (bottom) boundary.

As a possible choice of NH perturbations that preserve locality and pseudo-inversion symmetry with $\mathcal{I} =\tau_y \sigma_x$, we consider
\begin{align}
    &H (\bs{k}) = v \left( k_x \tau_z \sigma_x + k_y \tau_0 \sigma_y + k_z \tau_0 \sigma_z \right) \nonumber \\
    &\qquad\qquad\qquad\qquad\qquad+ \delta \tau_0 \sigma_x + \ii v' k_x \tau_0 \sigma_0.
\end{align}
These perturbations shift the Weyl points and open a point gap around $E = 0$, akin to the 2D NH Dirac surface modes considered in Eq.~\eqref{eq: 3D NH Ham AIII}.
In general, 3D NH systems are characterized by the 3D winding number $W_3$ of the complex spectrum~\cite{Kawabata2019}.
Around the reference energy $E=0$, the above NH Weyl Hamiltonian indeed realizes $W_3 \neq 0$, which leads to the emergence of the anomalous boundary modes~\cite{TerrierKunst_2020, eti, Kawabata2021, Nakamura2022}
\begin{align}
    E \left( \bs{k} \right) = \ii k_x + k_z
        \label{eq: EP-dispersion}
\end{align}
under OBC along the $y$-direction.
This complex-valued energy dispersion coincides with that of an exceptional point.
Importantly, if we perform a dimensional reduction by setting $k_z = 0$ in the 4D Chern insulator and 3D NH Weyl modes, and further impose chiral symmetry in Eq.~(\ref{eq: CS-NH}), the complex-valued dispersion in Eq.~(\ref{eq: EP-dispersion}) reduces to the imaginary-valued dispersion in Eq.~(\ref{eq: 3D AIII chiral hinge}). In this manner, the higher-order boundary modes protected by the boundary point-gap topology in different spatial dimensions are related to each other via dimensional reduction. While such a dimensional hierarchy is similar to that in Hermitian topological phases~\cite{Schnyder-Ryu-08, Kitaev-09, Ryu-10, Qi2008, Hasan-Kane-review, Qi-Zhang-review, Schnyder-Ryu-review}, the interplay of point gaps and line gaps, as well as the bifurcation of symmetry, enriches the boundary physics of NH topological phases.

\section{3D \texorpdfstring{$\mathbb{Z}_2$}{Z2}-classified topological insulator (class AII)}
\label{sec: 3DZ2}

3D Hermitian systems can generally host $\mathbb{Z}_2$ topological phases in classes AII and CII, in which time-reversal symmetry in Eq.~(\ref{eq: TRS}) with $\mathcal{T}^2 = -1$ is respected.
Similarly to the $\mathbb{Z}$ topological phase protected by chiral symmetry in Sec.~\ref{sec: 3DTI_story}, Dirac surface modes appear on each boundary for OBC along the $z$-direction, so that the low-energy Hamiltonian is Eq.~(\ref{eq: 2D chiral Dirac}) with time-reversal symmetry $\mathcal{T} = \mathrm{i} \tau_0 \sigma_y \mathit{K}$.
In contrast to the previous case, the Dirac point is fixed at the time-reversal-invariant momentum and cannot be shifted by symmetry-preserving perturbations, consistent with the $\mathbb{Z}_2$ topological phase.
In particular, the perturbation $\delta \tau_0 \sigma_x$ in Eq.~(\ref{eq: 3D NH Ham AIII}) breaks time-reversal symmetry and is forbidden in the presence of time-reversal symmetry.

Note that no point-gap topology is well defined in 2D class AII, in contrast to 2D class AII$^{\dag}$~\cite{Kawabata2019}. We therefore consider NH perturbations in class AII$^\dagger$, in which time-reversal symmetry$^{\dag}$ in Eq.~(\ref{eq: TRS-dag}) is respected.
The generic perturbed NH Hamiltonian in class AII$^\dagger$ that preserves locality and pseudo-inversion symmetry reads
\begin{equation}
    H(\bs{k}) = v k_x \tau_z \sigma_x + v k_y \tau_0 \sigma_y + \mathrm{i} \gamma \tau_z \sigma_0
        \label{eq: Dirac 2D AIIdag}
\end{equation}
with $\gamma \in \mathbb{R}$. 
The NH perturbation $\ii \gamma \tau_z \sigma_0$ opens a point gap for $E=0$, 
around which the $\mathbb{Z}_2$ topological invariant takes a nontrivial value, $\nu(0) = 1$~\cite{Kawabata2019}. Since this is the only symmetry-allowed constant NH perturbation, the 2D $\mathbb{Z}_2$ topological invariant $\nu(E)$ of the Dirac surface states in the slab geometry (OBC only in one direction) satisfies
\begin{equation}
   \nu(E_\mathrm{F}) = \nu
\end{equation}
for a NH system in class AII$^\dagger$ that is point-gapped at $E_\mathrm{F}$ and preserves pseudo-inversion symmetry (Tab.~\ref{tab: mapping3D}).

Importantly, this $\mathbb{Z}_2$ topological invariant is intrinsic to NH systems and leads to the skin effect, in contrast to the $\mathbb{Z}$ topological invariant protected by chiral symmetry that has a counterpart in anti-Hermitian systems and hence leads to the emergence of chiral hinge modes.
The skin effect is unique to 2D systems and can be probed by adding a magnetic flux~\cite{Okuma2020_toposkin, DennerSchindler2022_flux}.
Here, when we reduce the spatial dimension along the $y$-direction, the NH Dirac surface modes in Eq.~(\ref{eq: Dirac 2D AIIdag}) reduce to the NH helical edge modes in Eq.~(\ref{eq: Dirac 1D AIIdag}).
Correspondingly, the skin effect can also be understood by dimensional reduction.

\section{General classification}
While we have so far studied some prototypical models, our discussions can be generalized to arbitrary symmetry classes and spatial dimensions.
For Hermitian systems in each dimension, $\mathbb{Z}$ and $\mathbb{Z}_2$ topological phases appear in 5 symmetry classes out of the 10 AZ symmetry classes~\cite{Schnyder-Ryu-08, Kitaev-09, Ryu-10, Hasan-Kane-review, Qi-Zhang-review, Schnyder-Ryu-review}.
As also discussed before, each AZ symmetry in Hermitian systems has different generalizations in NH systems, which culminate in the 38-fold symmetry classification~\cite{Bernard-LeClair-02, Kawabata2019, Zhou-Lee-19}.
Notably, boundary modes in Hermitian topological insulators and superconductors always exhibit nontrivial point-gap topology in the presence of NH perturbations in the AZ$^{\dag}$ symmetry classes, in which time-reversal and chiral symmetries are defined by Eqs.~(\ref{eq: TRS-dag}) and (\ref{eq: CS-NH}), respectively.
This general correspondence exists because the classification of point-gap topology in $\left( d-1 \right)$ dimensions 
for an AZ$^{\dag}$ symmetry class
coincides with that of Hermitian topology in $d$ dimensions
for the corresponding AZ symmetry class~\cite{Kawabata2019}. This relationship was previously applied to dynamical phenomena of NH systems~\cite{Vishwanath_2019, Bessho2021}.
On the other hand, in the presence of NH perturbations in the AZ symmetry classes, in which time-reversal and chiral symmetries are respectively defined by Eqs.~(\ref{eq: TRS}) and (\ref{eq: CS-NH}), 3 classes out of the 5 classes can exhibit nontrivial point-gap topology.

Here, using continuum Dirac Hamiltonians, we explicitly derive the mapping between the Hermitian bulk topology and NH boundary topology in 2D and 3D for all the remaining AZ symmetry classes, as summarized in Tabs.~\ref{tab: mapping2D} and~\ref{tab: mapping3D}.
Some symmetry classes exhibit NH topology different from the models studied in Secs.~\ref{sec: Chern_story}, \ref{sec: 2DTI_story}, \ref{sec: 3DTI_story}, and \ref{sec: 3DZ2}.
For example, in 2D class D and 3D class DIII, no point gap is allowed to be open because of the Hermitian topology in the bulk.
In addition, in 3D class CII$^{\dag}$, NH boundary topology can lead to the emergence of helical hinge modes protected by point-gap topology.
It is also notable that the classification in arbitrary dimensions is related to each other via dimensional reduction; the generalization to higher dimensions is also straightforward.

\subsection{2D Class D} \label{sec: 2DclassD}
The topological invariant of 2D Hermitian systems in class D is the Chern number $C \in \mathbb{Z}$, similar to 2D class A (see Sec.~\ref{sec: Chern_story}). For the simplest nontrivial case $C = 1$, the low-energy continuum Hamiltonian in the slab geometry is therefore given by
\begin{equation} \label{eq: pplusip_edgemodes}
H_0(k_x) = v k_x \tau_z,
\end{equation}
which satisfies particle-hole symmetry:
\begin{equation}
\mathcal{P} H_0(\bs{k}) \mathcal{P}^\dagger = -H_0(-\bs{k}),
    \label{eq: PHS}
\end{equation}
with $\mathcal{P} = \tau_0 \mathit{K}$.
We choose $\tau_0$ rather than $\tau_z$ in the definition of $\mathcal{P}$ because particle-hole symmetry is an internal symmetry that is unaffected by spatial transformations such as those that exchange the two edges.

\subsubsection{NH class D}
NH systems in class D are defined to respect 
\begin{equation}
\mathcal{P} H_0(\bs{k}) \mathcal{P}^\dagger = -H^\dagger_0(-\bs{k}),
    \label{eq: PHS_NH}
\end{equation}
with an antiunitary operation $\mathcal{P}$ that satisfies $\mathcal{P}^2 =1$. Notably, Eq.~\eqref{eq: pplusip_edgemodes}, which describes one chiral mode per edge, cannot become real-line-gapped under any perturbations that preserve locality. In fact, for the case with $C=1$, even a point gap cannot open. 
Keeping the definition $\mathcal{P} = \tau_0 \mathit{K}$, the only NH mass term for Eq.~\eqref{eq: pplusip_edgemodes} that satisfies Eq.~\eqref{eq: PHS_NH} is $\mathrm{i} m \tau_y$, which couples the two edges and is therefore nonlocal.

For $C=2$, on the other hand, two pairs of chiral edge modes appear in the slab geometry, which read
\begin{align}  
    H_0(k_x) = v k_x \tau_z\sigma_0.
\end{align}
Here, the Pauli matrices $\sigma_i$ describe the two flavors of chiral modes at each edge.
In this case, the generic NH perturbations that preserve particle-hole symmetry, pseudo-inversion symmetry, and locality are
\begin{align}
    H(k_x) = v k_x \tau_z\sigma_0 + \ii b \tau_z \sigma_y
\end{align}
with $b \in \mathbb{R}$.
The complex spectrum is obtained as
\begin{align}  
    E(k_x) = \tau \left( v k_x + \ii b \sigma \right),
\end{align}
where $\tau = +1$ ($\tau = -1$) denotes the chiral edge modes at the top (bottom) boundary, and $\sigma = +1$ ($\sigma = -1$) denotes the different flavors of the chiral edge modes at each boundary.
Notably, a point gap is open around $E=0$ for $b \neq 0$, which contrasts with the previous case for $C=1$. 
This should be related to the $\mathbb{Z}_2$ classification of both point-gap and line-gap topology in 1D class D~\cite{Kawabata2019}.

\subsubsection{NH class \texorpdfstring{D$^\dagger$}{D†}}
We now consider NH perturbations in class D$^\dagger$. This symmetry class is characterized by the particle-hole constraint in Eq.~\eqref{eq: PHS}, and the allowed NH perturbations are
\begin{equation}
    H(k_x) = \mathrm{i} a \tau_0 + (v k_x + \mathrm{i} b) \tau_z,
\end{equation}
with $a,b \in \mathbb{R}$. 
In contrast to class D, a point gap is open in class D$^\dagger$.
Again, pseudo-inversion symmetry $\mathcal{I} = \tau_x$ results in $a = 0$ and protects a nonzero winding number $|W(0)| = 1$ for $b \neq 0$. In a similar manner to 2D class A in Sec.~\ref{sec: Chern_story}, we find that the 1D winding number $W(E)$ of the chiral edge states in the slab geometry (OBC only in one direction) satisfies
\begin{equation}
    W(0) = C \quad (\mathrm{mod}\ 2),
\end{equation}
this time for a NH system in class D$^\dagger$ that is point-gapped at $E = 0$ and preserves pseudo-inversion symmetry.

\subsection{2D Class DIII}
We have seen in Sec.~\ref{sec: 2DclassD} that the addition of particle-hole symmetry $\mathcal{P}$ to class A, which results in Hermitian class D and NH class D$^\dagger$, does not change the derivation of the boundary spectral winding in Sec.~\ref{sec: Chern_story}.
Similarly, the derivation in Sec.~\ref{sec: 2DTI_story} is fully compatible with additional particle-hole symmetry $\mathcal{P} = \tau_0 \sigma_0 \mathit{K}$ that upgrades Hermitian class AII to Hermitian class DIII and NH class AII$^\dagger$ to NH class DIII$^\dagger$, respectively. 
As a consequence, we again find that the 1D $\mathbb{Z}_2$ topological invariant $\nu(E)$ of the helical edge states in the slab geometry satisfies
\begin{equation}
    \nu(0) = \nu
\end{equation}
this time for a NH system in class DIII$^\dagger$ that is point-gapped at $E=0$ and preserves pseudo-inversion symmetry.

\subsection{2D Class C}
The minimal nontrivial topological phase of 2D Hermitian systems in class C is an insulator or a gapped superconductor with the Chern number $C = 2$, which has two co-propagating chiral edge modes in the slab geometry described by [cf. Eq.~\eqref{eq: chern2edgetheory}]
\begin{equation}
H_0(k_x) = v k_x \tau_z \sigma_0.
\end{equation}
This Hamiltonian satisfies particle-hole symmetry in Eq.~(\ref{eq: PHS}) with $\mathcal{P} = \tau_0 \sigma_y \mathit{K}$.
In contrast to class D, the particle-hole operation in class C satisfies $\mathcal{P}^2 = -1$.

We now consider NH perturbations in class C$^\dagger$, because this is the only NH generalization of Hermitian class C that allows for nontrivial point-gap topology in 1D~\cite{Kawabata2019}. 
This symmetry class is characterized by the particle-hole constraint in Eq.~\eqref{eq: PHS}, and the allowed NH perturbations are
\begin{equation}
    H(k_x) = \mathrm{i} a \tau_0 \sigma_0 + (v k_x + \mathrm{i} b) \tau_z \sigma_0,
\end{equation}
with $a,b \in \mathbb{R}$. Again, pseudo-inversion symmetry $\mathcal{I} = \tau_x \sigma_0$ results in $a = 0$ and protects the nonzero winding number $|W(0)| = 2$ for $b \neq 0$.

We next study a system with the Chern number $C = 4$. 
At low energies and in the slab geometry, the system hosts two pairs of the chiral edge states described by
\begin{equation}
H_0(k_x) = v k_x \tau_z \sigma_0 \rho_0,
\end{equation}
where $\rho_i$ ($i=0,x,y,z$) is another set of Pauli matrices that acts on the different flavors of the chiral modes at each edge. 
Particle-hole symmetry is now represented by $\mathcal{P} = \tau_0 \sigma_y \rho_0 \mathit{K}$. 
One example of NH perturbations in class C$^\dagger$ is given by
\begin{equation}
H_0(k_x) = v k_x \tau_z \sigma_0 \rho_0 + \mathrm{i} a \tau_z \sigma_0 \rho_0 + \mathrm{i} b \tau_z \sigma_0 \rho_z,
\end{equation}
with $a,b \in \mathbb{R}$. 
For $a > 0$ and $b = 0$, we now obtain $W(0) = 4$, while for $a = 0$ and $b > 0$, we obtain $W(0) = 0$. Hence, pseudo-inversion symmetry allows us to fix the boundary winding number only modulo $4$.

As a consequence, the 1D NH winding number $W(E)$ of the pairs of the chiral edge modes in the slab geometry (OBC only in one direction) satisfies
\begin{equation}
    W(0) = C \quad (\mathrm{mod}\ 4)
\end{equation}
for a NH system in class C$^\dagger$ that is point-gapped at $E=0$ and preserves pseudo-inversion symmetry.

\subsection{3D class DIII}
Similarly to class AIII, the topological invariant of 3D Hermitian systems in class DIII is the 3D winding number $W \in \mathbb{Z}$.
Correspondingly, for the simplest nontrivial case $W=1$, the low-energy continuum Hamiltonian in the slab geometry is given as the Dirac surface modes in Eq.~(\ref{eq: 2D chiral Dirac}),
which respect time-reversal symmetry in Eq.~(\ref{eq: TRS}) with $\mathcal{T} = \ii \tau_0 \sigma_y K$ and chiral symmetry in Eq.~(\ref{eq: CS}) with $\CS = \tau_0 \sigma_z$.
Class DIII in Hermitian systems can be generalized to NH systems in two manners, classes DIII and DIII$^{\dag}$~\cite{Kawabata2019}.
As shown below, while the topological invariant $W$ in the 3D bulk leads to point-gap ingappability in class DIII, it induces a skin effect or the emergence of chiral hinge modes in class DIII$^{\dag}$.

\subsubsection{NH class DIII}
NH systems in class DIII are defined to respect time-reversal symmetry in Eq.~(\ref{eq: TRS}) and chiral symmetry in Eq.~(\ref{eq: CS-NH}). Similar to the NH chiral edge modes in 2D class D, no point gap can open in the NH Dirac surface modes in class DIII with $W = 1$.
More specifically, the generic NH Dirac surface modes that preserve time-reversal and chiral symmetries, as well as locality and pseudo-inversion symmetry with $\mathcal{I}=\tau_y \sigma_x$, are of the form
\begin{align}
    H(\bs{k}) = v k_x \tau_z \sigma_x + v k_y \tau_0 \sigma_y + 
    \ii \gamma \tau_0 \sigma_z
\end{align}
with $\gamma \in \mathbb{R}$. The spectrum is obtained as
\begin{equation}
E(\bs{k}) = \pm \sqrt{v^2 k_x^2 + v^2 k_y^2 - \gamma^2},
\end{equation}
which exhibits a loop of exceptional points at $k_x^2 + k_y^2 = \left( \gamma/v \right)^2$ in momentum space and possesses no point or line gaps.

\subsubsection{NH class \texorpdfstring{DIII$^\dagger$}{DIII†}}

In class DIII$^{\dag}$, NH systems are defined to respect time-reversal symmetry$^{\dag}$ in Eq.~(\ref{eq: TRS-dag}) and chiral symmetry in Eq.~(\ref{eq: CS-NH}).
In general, such systems are characterized by a $\mathbb{Z}$ topological invariant [i.e., Chern number $C(0)$ of the Hermitian Hamiltonian $\ii H(\bs{k})\CS$ evaluated for the reference energy $E=0$] in the presence of a point gap~\cite{Kawabata2019}.
If $C(0)$ is odd, the skin effect occurs~\cite{Okuma2020_toposkin, Nakamura2022}. 
On the other hand, even $C(0)$ does not lead to the skin effect but the appearance of chiral hinge modes.
Consequently, the surface Dirac cone in Eq.~(\ref{eq: 2D chiral Dirac}) in class DIII$^{\dag}$, which corresponds to $C(0)=\pm1$, only exhibits the skin effect rather than chiral hinge modes.
Specifically, the only NH perturbation in class DIII$^{\dag}$ preserving pseudo-inversion symmetry with $\mathcal{I} = \tau_y\sigma_x$ is
\begin{equation}
    H(\bs{k}) = v k_x \tau_z \sigma_x + v k_y \tau_0 \sigma_y + \ii \gamma \tau_z \sigma_0,
\end{equation}
with $\gamma \in \mathbb{R}$. Notably, this NH Dirac surface Hamiltonian has a net nonzero Chern number of $C(0) = 1$. This leads to the skin effect that is unique to 2D systems and can be probed by a magnetic flux~\cite{Okuma2020_toposkin, DennerSchindler2022_flux}.

On the other hand, when the original 3D topological insulator exhibits the 3D winding number $W=2$, two different flavors of the Dirac surface modes appear at each boundary. These Dirac surface modes could in principle yield a Chern number of $C(0)=2$ protected by chiral symmetry, which would result in two flavors of chiral hinge modes under OBC. However, such an even number of the Chern number cannot be fixed solely by pseudo-inversion symmetry---there are other symmetry-preserving perturbations that lead to $C(0)=0$---so that we can only deduce 
\begin{equation}
C(0) = W \quad (\text{mod } 2)
\end{equation}
in general.

\subsection{3D class CI}
The minimal nontrivial topological phase of 3D Hermitian systems in class CI is a gapped superconductor with the winding number $W = 2$ that hosts two Dirac surface modes, so that the low-energy slab Hamiltonian reads
\begin{equation} \label{eq: ci_surface_unperturbed_theory}
H_0(\bs{k}) = v \left( k_x \tau_z \sigma_x + k_y \tau_0 \sigma_y \right) \rho_0.
\end{equation}
This Dirac Hamiltonian satisfies particle-hole symmetry in Eq.~(\ref{eq: PHS}) with $\mathcal{P} = \ii \tau_0 \sigma_x \rho_y \mathit{K}$ and chiral symmetry in Eq.~(\ref{eq: CS}) with $\CS = \tau_0 \sigma_z \rho_0$.
Notably, this is a pair of the Dirac surface modes in Eq.~(\ref{eq: 2D chiral Dirac}) with the same chirality.

We now consider NH perturbations in class CI$^\dagger$, because this is the only NH generalization of class CI that allows for nontrivial point-gap topology in 2D~\cite{Kawabata2019}. 
One possible choice of NH perturbations that preserve time-reversal and chiral symmetries, as well as locality and pseudo-inversion symmetry with $\mathcal{I} = \tau_y \sigma_x \rho_0$, is 
\begin{align} \label{eq: ci_surface_perturbed_theory}
    &H(\bs{k}) = v \left( k_x \tau_z \sigma_x + k_y \tau_0 \sigma_y \right) \rho_0  \nonumber \\
    &\qquad\qquad\qquad + \tau_0 \left( \delta \sigma_x + \ii v' k_x \sigma_0 \right) \rho_z
\end{align}
with $\delta, v' \in \mathbb{R}$. 
This NH Dirac surface Hamiltonian possesses a point gap around $E=0$, for which the Chern number of $\ii H(\bs{k}) \CS$ is nontrivial, i.e., $|C(0)| = 2$.
As a consequence, it hosts the two flavors of chiral hinge modes with the imaginary-valued dispersion in Eq.~(\ref{eq: 3D AIII chiral hinge}).
Moreover, the only local and $\bs{k}$-independent NH perturbations to Eq.~\eqref{eq: ci_surface_unperturbed_theory} allowed by symmetry are of the form $+\mathrm{i} m_{1,i} \tau_0 \sigma_z \rho_i$ ($i=x,y,z$), which do not open a gap in $\ii H(\bs{k}) \CS$ (i.e., they are not Dirac mass terms), and $+\mathrm{i} m_{2} \tau_z \sigma_0 \rho_0$, which again results in $|C(0)| = 2$. In conclusion, the 2D Chern number $C(0)$ of the NH Dirac surface modes in the slab geometry (OBC only in one direction) must satisfy
\begin{equation}
    C(0) = W \quad (\mathrm{mod}\ 4)
\end{equation}
for a NH system in class CI$^\dagger$ that is point-gapped at $E=0$ and preserves pseudo-inversion symmetry.

\subsection{3D class CII}
The nontrivial topological phase of 3D Hermitian systems in class CII is characterized by a $\mathbb{Z}_2$ invariant $\nu=1$. The low-energy Dirac Hamiltonian at the boundaries of such a system is
\begin{align} \label{eq: cii_surface_pristine} 
    H_0(\bs{k}) = v \left( k_x \tau_z \sigma_x + k_y \tau_0 \sigma_y \right) \rho_x, 
\end{align}
which respects chiral symmetry in Eq.~(\ref{eq: CS}) with $\CS = \tau_0\sigma_0\rho_z$ and time-reversal symmetry in Eq.~(\ref{eq: TRS}) with $\mathcal{T} = \ii \tau_0 \sigma_y \rho_0 K$. 
While 2D NH point-gapped systems in class CII are classified by a $2\mathbb{Z}$ topological invariant, those in class CII$^{\dag}$ are classified only by a $\mathbb{Z}_2$ topological invariant.
Owing to this difference, an even number of chiral hinge modes appear in class CII, while helical hinge modes appear in class CII$^{\dag}$, as shown below.

\subsubsection{NH class CII}
We first consider NH perturbations in class CII, which are defined to respect time-reversal symmetry in Eq.~(\ref{eq: TRS}) and chiral symmetry in Eq.~(\ref{eq: CS-NH}).
One possible choice of NH perturbations that preserve time-reversal and chiral symmetries, as well as locality and pseudo-inversion symmetry with $\mathcal{I} = \tau_y \sigma_x \rho_x$, is 
\begin{align} \label{eq: cii_surface_perturbations} 
    &H(\bs{k}) = v \left( k_x \tau_z \sigma_x + k_y \tau_0 \sigma_y \right) \rho_x  \nonumber \\
    &\qquad\qquad\qquad + \tau_0 \left( \delta \sigma_y \rho_y + \ii v' k_x \sigma_0 \rho_0 \right)
\end{align}
with $\delta, v' \in \mathbb{R}$. 
This NH Dirac surface Hamiltonian possesses a point gap for $E=0$, around which the Chern number of $\ii H(\bs{k}) \CS$ is nontrivial, i.e., $|C(0)| = 2$.
As a consequence, it hosts two flavors of chiral hinge modes with the imaginary-valued dispersion in Eq.~(\ref{eq: 3D AIII chiral hinge}).

Moreover, the only local and $\bs{k}$-independent NH perturbations to Eq.~\eqref{eq: cii_surface_pristine} allowed by symmetry are of the form $\mathrm{i} m_1 \tau_0 \sigma_x \rho_z$, $\mathrm{i} m_2 \tau_0 \sigma_y \rho_0$, $\mathrm{i} m_3 \tau_0 \sigma_z \rho_0$, $\mathrm{i} m_4 \tau_z \sigma_x \rho_0$, $\mathrm{i} m_5 \tau_z \sigma_y \rho_z$, $\mathrm{i} m_6 \tau_z \sigma_z \rho_z$. 
Of these, only the last serves as a Dirac mass in $\ii H(\bs{k}) \CS$ and leads to $C(0) = \pm 2$, and the remaining terms do not open a gap. In conclusion, the 2D Chern number $C(0)$ of the NH Dirac surface modes in the slab geometry (OBC only in one direction) must satisfy
\begin{equation}
    C(0) = 2\nu \quad (\mathrm{mod}\ 4)
\end{equation}
for a NH system in class CII that is point-gapped at $E=0$ and preserves pseudo-inversion symmetry.

\subsubsection{NH class \texorpdfstring{CII$^\dagger$}{CII†}}

We next consider NH perturbations in class CII$^{\dag}$, which are defined to respect time-reversal symmetry$^{\dag}$ in Eq.~(\ref{eq: TRS-dag}) and chiral symmetry in Eq.~(\ref{eq: CS-NH}).
Because of time-reversal symmetry$^{\dag}$ for the NH Hamiltonian $H(\bs{k})$, the Hermitian matrix $\ii H(\bs{k}) \CS$ also respects time-reversal symmetry and hence must have the vanishing Chern number.
Instead, $\ii H(\bs{k}) \CS$ hosts the well-defined $\mathbb{Z}_2$ topological invariant protected by time-reversal symmetry$^{\dag}$~\cite{Kawabata2019}.
In fact, one possible choice of NH perturbations in class CII$^{\dag}$ for $\mathcal{I} = \tau_y \sigma_x \rho_x$ reads 
\begin{align}
    &H(\bs{k}) = v \left( k_x \tau_z \sigma_x + k_y \tau_0 \sigma_y \right) \rho_x  \nonumber \\
    &\qquad\qquad\qquad + \tau_0 \left( \delta \sigma_y \rho_y + \ii v' k_x \sigma_z \rho_z \right).
\end{align}
This NH Dirac surface Hamiltonian possesses a point gap around $E=0$, and we have
\begin{align}
    &\ii H(\bs{k}) \CS = v k_x \tau_z \sigma_x \rho_y + v k_y \tau_0 \sigma_y \rho_y \nonumber \\
    &\qquad\qquad\qquad\quad- \delta \tau_0 \sigma_y \rho_x - v' k_x \tau_0 \sigma_z \rho_0,
\end{align}
which exhibits the nontrivial $\mathbb{Z}_2$ invariant $\nu(0) = 1$ evaluated for the reference energy $E=0$ as fixed by chiral symmetry.
Correspondingly, under OBC along both $y$ and $z$-directions, the NH Dirac surface modes host the helical hinge modes
\begin{align}
    H (k_x) = \ii v' k_x \tau_0 \sigma_z.
\end{align}
Moreover, the only local and $\bs{k}$-independent NH perturbations to Eq.~\eqref{eq: cii_surface_pristine} allowed by symmetry are now of the form $\mathrm{i} m_1 \tau_0 \sigma_0 \rho_z$ and $\mathrm{i} m_2 \tau_z \sigma_0 \rho_0$. Of these, the first one does not open a gap in $\ii H(\bs{k}) \CS$ while the second term again results in $\nu(0) = 1$. In conclusion, the 2D $\mathbb{Z}_2$ topological invariant $\nu(0)$ of the NH Dirac surface modes in the slab geometry (OBC only in one direction) must satisfy
\begin{equation}
    \nu(0) = \nu
\end{equation}
for a NH system in class CII$^{\dag}$ that is point-gapped at $E=0$ and preserves pseudo-inversion symmetry.

\section{Discussion}
Perturbing a gapped Hermitian system with small NH terms generally results in a line-gapped spectrum that can be adiabatically deformed back to the Hermitian limit. To realize intrinsic NH topology, one must therefore perturb a gapless system or make sure that the NH terms are significantly larger than the gap energy scale. We have found that Hermitian topological phases allow us to avoid such fine tuning because they are guaranteed to host gapless states on their boundaries. Moreover, we have shown that the resulting boundary NH topological invariants are determined by the bulk Hermitian invariants in the presence of pseudo-inversion symmetry.

Besides generalizing our framework to other spatial symmetries beyond pseudo-inversion symmetry, a natural follow-up question is to consider the effect of NH perturbations on the gapless boundary states of topological crystalline insulators. For instance, NH perturbations should transform the gapless hinge states of second-order 3D topological insulators and superconductors into third-order skin effects appearing at the corners of samples terminated in all three directions. It is also worthwhile to study NH boundary topology in Floquet topological insulators~\cite{LiuFulga2022}.

Another natural question is to investigate boundary NH topology from a field-theoretic perspective. 
For instance, each edge of a Chern insulator in the presence of NH perturbations (Sec.~\ref{sec: Chern_story}) realizes \emph{half} of a Hatano-Nelson chain~\cite{Hatano1996,Hatano1997} and cannot be regularized in a 1D lattice system. Here, the presence of the chiral edge states is supported by the quantum anomaly of the Hermitian bulk system in 2D.
On the other hand, the complex-spectral winding number and the concomitant skin effect are relevant to another quantum anomaly that is unique to NH systems and formulated in terms of spatial, rather than spatiotemporal, degrees of freedom~\cite{Kawabata2021}.
It is of interest to study the interplay of these two types of quantum anomaly.

After the initial submission of this work, the corner skin effect of chiral edge modes in a NH Chern insulator was experimentally observed in a lossy gyromagnetic photonic crystal~\cite{Liu2023exp}. 
Other NH models in our work, such as the 3D model studied in Sec.~\ref{sec: 3DTI_story}, should also be realizable in similar open classical and quantum synthetic materials.
Furthermore, our work is also relevant to topological insulators and superconductors in solid-state materials.
In this respect, it is noteworthy that the NH topology and concomitant skin effect were recently demonstrated in the conductance matrix for the mesoscopic edge transport of a quantum Hall device~\cite{Ochkan2023exp}.

\medskip
{\it Note added.---} In the final stages of completion of this work, we became aware of a recent related work~\cite{Ma2023}. This related work investigates the skin effect of a chiral edge mode in a NH Chern insulator, which is consistent with our discussions in Sec.~\ref{sec: Chern_story}.
We also note another related work~\cite{Nakamura2023} that appeared after the submission of this work. Our results are consistent where they overlap.

\begin{acknowledgments}
F.S.~is supported by a fellowship at the Princeton Center for Theoretical Science. B.L.~is supported by the Alfred P. Sloan Foundation, the National Science Foundation through Princeton University’s Materials Research Science and Engineering Center DMR-2011750, and the National Science Foundation under award DMR-2141966. K.K.~is supported by the Japan Society for the Promotion of Science (JSPS) through the Overseas Research Fellowship, and the Gordon and Betty Moore Foundation through Grant No.~GBMF8685 toward the Princeton theory program. This research was supported in part by the National Science Foundation under Grant No. NSF PHY-1748958.
\end{acknowledgments}

\appendix

\section{Relation between Chern higher-order NH skin effect and Hermitian fragile topology}
    \label{appendix: fragile}

We here give an equivalent proof of the higher-order skin effect in NH Chern insulators that only relies on bulk topology and does not make use of the slab geometry. Consider a Chern insulator with $C \neq 0$ (mod $2$) that is gapped at the Fermi energy $E_\mathrm{F}$. We now show that taking into account pseudo-inversion symmetry
\begin{equation}
    \mathcal{I} H(\bs{k}) \mathcal{I}^\dagger = H(-\bs{k})^\dagger, \quad \mathcal{I}^2 = 1,
\end{equation}
implies that a second-order skin effect \emph{must} occur as long as the point gap at $E_\mathrm{F}$ is open. For this, we note that the extended Hermitian Hamiltonian
\begin{equation}
    \bar{H}(\bs{k}) = \begin{pmatrix} 0 & H(\bs{k}) \\ H(\bs{k})^\dagger & 0 \end{pmatrix}
\end{equation}
inherits conventional inversion symmetry
\begin{equation}
    \bar{\mathcal{I}} = \begin{pmatrix} 0 & \mathcal{I} \\ \mathcal{I} & 0 \end{pmatrix}, \quad
    \bar{\mathcal{I}} \bar{H}(\bs{k}) \bar{\mathcal{I}}^\dagger = \bar{H}(-\bs{k}).
\end{equation}
Moreover, $\bar{H}(\bs{k})$ always enjoys chiral (sublattice) symmetry
\begin{equation} \label{eq: chiralsymEHH}
    \bar{\Sigma} = \begin{pmatrix} \mathbb{1} & 0 \\ 0 & -\mathbb{1} \end{pmatrix}, \quad
    \bar{\Sigma} \bar{H}(\bs{k}) \bar{\Sigma}^\dagger = -\bar{H}(\bs{k}).
\end{equation}
by construction.

$\bar{\mathcal{I}}$ symmetry gives rise to a $\mathbb{Z}_4$-valued symmetry indicator topological invariant~\cite{Khalaf2018}
\begin{equation}
    \bar{\kappa} = \left[2 \bar{n}_{\mathrm{occ}} + \frac{1}{2} \sum_{\bs{k} \in \mathrm{HSMs}} \sum_{n \in \mathrm{occ}} \bar{\lambda}_n (\bs{k}) \right]\quad(\mathrm{mod}~4),
\end{equation}
where $\bar{n}_{\mathrm{occ}}$ is the number of occupied bands, $\mathrm{HSMs}$ are the four inversion-symmetric (``high-symmetry") momenta of the 2D Brillouin zone, and $\bar{\lambda}_n (\bs{k})$ are the $\bar{\mathcal{I}}$-eigenvalues of the occupied (negative-energy) eigenstates of $\bar{H}(\bs{k})$. All trivial (atomic insulator) band structures satisfy $\bar{\kappa} = 0$. It is also known that the Chern number $\bar{C}$ of the \emph{extended} Hamiltonian $\bar{H}(\bs{k})$ satisfies
\begin{equation} \label{eq: c_kappa_rel}
    \bar{C} = \bar{\kappa} \quad (\mathrm{mod}\ 2),
\end{equation}
so that the odd values $\bar{\kappa} = 1,3$ must correspond to Chern insulators.
Moreover, if an insulator has $\bar{\kappa} = 2$ but $\bar{C} = 0$, it must belong to an inversion-protected \emph{fragile topological phase}, which hosts $2+4l$ ($l \in \mathbb{Z}$) corner states for full OBC~\cite{Ahn19,Schindler2022}. Due to $\bar{\mathcal{I}}$ symmetry, they appear at opposite corners, and due to chiral symmetry [Eq.~\eqref{eq: chiralsymEHH}], they are pinned to zero energy.

We now compute $\bar{\kappa}$ in the situation where $H(\bs{k})$ [but not $\bar{H}(\bs{k})$] realizes a Chern insulator with $C \neq 0$ (mod $2$). For simplicity, we first adiabatically deform $H(\bs{k})$ to its Hermitian limit $H(\bs{k})^\dagger = H(\bs{k})$. We may then rotate to a basis where
\begin{equation}
    \bar{H}(\bs{k}) = \begin{pmatrix} H(\bs{k}) & 0 \\ 0 & -H(\bs{k}) \end{pmatrix}, \quad 
    \bar{\mathcal{I}} = \begin{pmatrix} \mathcal{I} & 0 \\ 0 & -\mathcal{I} \end{pmatrix}.
\end{equation}
We see that the occupied subspace of $\bar{H}(\bs{k})$ is formed by the ``occupied" subspace of $H(\bs{k})$ (the states below $E_\mathrm{F} = 0$), together with the ``empty" subspace of $H(\bs{k})$ (the states above $E_\mathrm{F} = 0$). Importantly, for the latter set of states, the sign of inversion is flipped. Since the sum of occupied and empty subspaces of any lattice model has vanishing Chern number, we deduce $\bar{C} = 0$. Moreover, we compute
\begin{equation}
    \bar{\kappa} = 2 \kappa = 2 \quad \left( \mathrm{mod}~4 \right),
\end{equation}
where we use the fact that the symmetry indicator $\kappa$ of $H(\bs{k})$ must satisfy $\kappa = 1$ (mod $2$) due to Eq.~\eqref{eq: c_kappa_rel}. 

Correspondingly, a  model $H(\bs{k})$ with $C \neq 0$ (mod $2$) and in the presence of pseudo-inversion symmetry has an extended Hermitian Hamiltonian $\bar{H}(\bs{k})$ that realizes an inversion-symmetry-protected fragile phase with two stable corner-localized zero modes. Since zero modes in $\bar{H}(\bs{k})$ induce a skin effect in $H(\bs{k})$~\cite{Okuma2020_toposkin}, we deduce that $H(\bs{k})$ has a \emph{higher-order} skin effect due to its first-order boundary-induced point gap topology. It is worthwhile to further study the relevance of this discussion to the second-order skin effects in Refs.~\onlinecite{Lee2019, Okugawa2020, Kawabata2020}.

\bibliography{refs}

\end{document}